\documentclass[a4paper]{spie} 
\usepackage[]{graphicx}
\usepackage{amssymb}

\newcommand{\lat}{Fermi-LAT}
\newcommand{\chandra}{Chandra}
\newcommand{\xmm}{XMM-Newton}
\newcommand{\suzaku}{Suzaku}
\newcommand{\integral}{INTEGRAL}

\newcommand{\nustar}{NuSTAR}
\newcommand{\astroh}{ASTRO-H}

\title{The  ASTRO-H X-ray Observatory} 

\normalsize
 \normalsize
\author{Tadayuki Takahashi\supit{a},
Kazuhisa~Mitsuda\supit{a},
Richard~Kelley\supit{b},
Henri~Aarts\supit{c},\\
Felix~Aharonian\supit{d},
Hiroki~Akamatsu\supit{c},
Fumie~Akimoto\supit{e},
Steve~Allen\supit{f},
Naohisa~Anabuki\supit{g},
Lorella~Angelini\supit{b},
Keith~Arnaud\supit{h},
Makoto~Asai\supit{f},
Marc~Audard\supit{i},
Hisamitsu~Awaki\supit{j},
Philipp~Azzarello\supit{i},
Chris~Baluta\supit{a},
Aya~Bamba\supit{k},
Nobutaka~Bando\supit{a},
Mark~Bautz\supit{l},
Roger~Blandford\supit{f},
Kevin~Boyce\supit{b},
Greg~Brown\supit{m},
Ed~Cackett\supit{n},
Maria~Chernyakova\supit{d},
Paolo~Coppi\supit{o},
Elisa~Costantini\supit{c},
Jelle~de~Plaa\supit{c},
Jan-Willem~den~Herder\supit{c},
Michael~DiPirro\supit{b},
Chris~Done\supit{p},
Tadayasu~Dotani\supit{a},
John~Doty\supit{q},
Ken~Ebisawa\supit{a},
Megan~Eckart\supit{b},
Teruaki~Enoto\supit{r},
Yuichiro~Ezoe\supit{s},
Andrew~Fabian\supit{n},
Carlo~Ferrigno\supit{i},
Adam~Foster\supit{t},
Ryuichi~Fujimoto\supit{u},
Yasushi~Fukazawa\supit{v},
Stefan~Funk\supit{f},
Akihiro~Furuzawa\supit{e},
Massimiliano~Galeazzi\supit{w},
Luigi~Gallo\supit{x},
Poshak~Gandhi\supit{a},
Keith~Gendreau\supit{b},
Kirk~Gilmore\supit{f},
Daniel~Haas\supit{c},
Yoshito~Haba\supit{e},
Kenji~Hamaguchi\supit{h},
Isamu~Hatsukade\supit{y},
Takayuki~Hayashi\supit{a},
Kiyoshi~Hayashida\supit{g},
Junko~Hiraga\supit{z},
Kazuyuki~Hirose\supit{a},
Ann~Hornschemeier\supit{b},
Akio~Hoshino\supit{u},
John~Hughes\supit{aa},
Una~Hwang\supit{ab},
Ryo~Iizuka\supit{ac},
Yoshiyuki~Inoue\supit{f},
Kazunori~Ishibashi\supit{e},
Manabu~Ishida\supit{a},
Kosei~Ishimura\supit{a},
Yoshitaka~Ishisaki\supit{s},
Masayuki~Ito\supit{ad},\\
Naoko~Iwata\supit{a},
Naoko~Iyomoto\supit{ae},
Jelle~Kaastra\supit{c},
Timothy~Kallman\supit{b},
Tuneyoshi~Kamae\supit{f},
Jun~Kataoka\supit{af},
Satoru~Katsuda\supit{r},
Hajime~Kawahara\supit{s},
Madoka~Kawaharada\supit{a},
Nobuyuki~Kawai\supit{ag},
Shigeo~Kawasaki\supit{a},
Dmitry~Khangaluyan\supit{a},
Caroline~Kilbourne\supit{b},
Masashi~Kimura\supit{g},
Kenzo~Kinugasa\supit{ah},
Shunji~Kitamoto\supit{ai},
Tetsu~Kitayama\supit{aj},
Takayoshi~Kohmura\supit{ak},
Motohide~Kokubun\supit{a},
Tatsuro~Kosaka\supit{al},
Alex~Koujelev\supit{am},
Katsuji~Koyama\supit{an},
Hans~Krimm\supit{b},
Aya~Kubota\supit{ao},
Hideyo~Kunieda\supit{e},
Stephanie~LaMassa\supit{o},
Philippe~Laurent\supit{ap},
Fran\c{c}ois~Lebrun\supit{ap},
Maurice~Leutenegger\supit{b},
Olivier~Limousin\supit{ap},
Michael~Loewenstein\supit{b},
Knox~Long\supit{aq},
David~Lumb\supit{ar},
Grzegorz~Madejski\supit{f},
Yoshitomo~Maeda\supit{a},
Kazuo~Makishima\supit{z},
Genevi\`eve~Marchand\supit{am},
Maxim~Markevitch\supit{b},
Hironori~Matsumoto\supit{e},
Kyoko~Matsushita\supit{as},
Dan~McCammon\supit{at},
Brian McNamara\supit{au},
Jon~Miller\supit{av},
Eric~Miller\supit{l},
Shin~Mineshige\supit{an},
Kenji~Minesugi\supit{a},
Ikuyuki~Mitsuishi\supit{s},
Takuya~Miyazawa\supit{e},
Tsunefumi~Mizuno\supit{v},
Hideyuki~Mori\supit{a},
Koji~Mori\supit{y},
Koji~Mukai\supit{b},
Toshio~Murakami\supit{u},
Hiroshi~Murakami\supit{ai},
Richard~Mushotzky\supit{h},
Housei~Nagano\supit{e},
Ryo~Nagino\supit{g},
Takao~Nakagawa\supit{a},
Hiroshi~Nakajima\supit{g},
Takeshi~Nakamori\supit{af},
Kazuhiro~Nakazawa\supit{z},
Yoshiharu~Namba\supit{aw},
Chikara~Natsukari\supit{a},
Yusuke~Nishioka\supit{y},
Masayoshi~Nobukawa\supit{an},
Masaharu~Nomachi\supit{g},
Steve~O'~Dell\supit{ax},
Hirokazu~Odaka\supit{a},
Hiroyuki~Ogawa\supit{a},
Mina~Ogawa\supit{a},
Keiji~Ogi\supit{j},
Takaya~Ohashi\supit{s},
Masanori~Ohno\supit{v},
Masayuki~Ohta\supit{a},
Takashi~Okajima\supit{b},
Atsushi~Okamoto\supit{ay},
Tsuyoshi~Okazaki\supit{a},
Naomi~Ota\supit{az},
Masanobu~Ozaki\supit{a},
Frits~Paerels\supit{ba},
St$\acute{\rm e}$phane~Paltani\supit{i},
Arvind~Parmar\supit{bb}, 
Robert~Petre\supit{b}, 
Martin~Pohl\supit{i},
F.Scott~Porter\supit{b},
Brian~Ramsey\supit{ax},
Rubens~Reis\supit{av},
Christopher~Reynolds\supit{h},
Helen~Russell\supit{au},
Samar~Safi-Harb\supit{bc},
Shin-ichiro~Sakai\supit{a},
Hiroaki~Sameshima\supit{a},
Jeremy~Sanders\supit{n},
Goro~Sato\supit{a},
Rie~Sato\supit{a},
Yoichi~Sato\supit{ay},
Kosuke~Sato\supit{as},
Makoto~Sawada\supit{k},
Peter~Serlemitsos\supit{b},
Hiromi~Seta\supit{ai},
Yasuko~Shibano\supit{a},
Maki~Shida\supit{a},
Takanobu~Shimada\supit{a},
Keisuke~Shinozaki\supit{ay},
Peter~Shirron\supit{b},
Aurora~Simionescu\supit{f},
Cynthia~Simmons\supit{b},
Randall~Smith\supit{t},
Gary~Sneiderman\supit{b},
Yang~Soong\supit{b},
Lukasz~Stawarz\supit{a},
Yasuharu~Sugawara\supit{ac},
Hiroyuki~Sugita\supit{ay},
Satoshi~Sugita\supit{e},
Andrew~Szymkowiak\supit{o},
Hiroyasu~Tajima\supit{e},
Hiromitsu~Takahashi\supit{v},
Shin-ichiro Takeda\supit{a},
Yoh~Takei\supit{a},
Toru~Tamagawa\supit{r},
Takayuki~Tamura\supit{a},
Keisuke~Tamura\supit{e},
Takaaki~Tanaka\supit{an},
Yasuo~Tanaka\supit{a},\\
Makoto~Tashiro\supit{bd},
Yuzuru~Tawara\supit{e},
Yukikatsu~Terada\supit{bd},
Yuichi~Terashima\supit{j},
Francesco~Tombesi\supit{b},
Hiroshi~Tomida\supit{a},
Yoko~Tsuboi\supit{ac},
Masahiro~Tsujimoto\supit{a},
Hiroshi~Tsunemi\supit{g},
Takeshi~Tsuru\supit{an},
Hiroyuki~Uchida\supit{an},
Yasunobu~Uchiyama\supit{f},
Hideki~Uchiyama\supit{z},
Yoshihiro~Ueda\supit{an},
Shiro~Ueno\supit{ay},
Shinichiro~Uno\supit{be},
Meg~Urry\supit{o},
Eugenio~Ursino\supit{w},
Cor~de~Vries\supit{c},
Atsushi~Wada\supit{ay},
Shin~Watanabe\supit{a},
Norbert~Werner\supit{f},
Nicholas~White\supit{b},
Takahiro~Yamada\supit{a},
Shinya~Yamada\supit{r},
Hiroya~Yamaguchi\supit{t},\\
Noriko~Yamasaki\supit{a},
Shigeo~Yamauchi\supit{az},
Makoto~Yamauchi\supit{y},
Yoichi~Yatsu\supit{ag},
Daisuke~Yonetoku\supit{u},
Atsumasa~Yoshida\supit{k}
Takayuki~Yuasa\supit{a}
\skiplinehalf
\supit{a}Institute of Space and Astronautical Science (ISAS), JAXA, Kanagawa 252-5210, Japan;\\
\supit{b}NASA/Goddard Space Flight Center, Greenbert, MD 20771, USA;\\
\supit{c}SRON Netherlands Institute for Space Research, Utrecht, the Netherlands;\\
\supit{d}Dublin Institute for Advanced Studies, Dublin, Ireland;
\supit{e}Department of Physics, Nagoya University, Nagoya 338-8570, Japan;
\supit{f}Kavli Institute for Particle Astrophysics and Cosmology, Stanford University, Stanford, CA 94305, USA;
\supit{g}Department of Earth and Space Science, Osaka University, Osaka 560-0043, Japan;
\supit{h}Department of Physics, University of Maryland, College Park, MD 21250, USA;
\supit{i}Universit\'e de Geneve24,  Geneve, Switzerland;
\supit{j}Department of Physics, Ehime University, Ehime 790-8577, Japan;\\
\supit{k}Department of Physics and Mathematics, Aoyama Gakuin University,  Kanagawa 229-8558, Japan;
\supit{l}Kavli Institute for Astrophysics and Space Research, Massachusetts Institute of Technology,  Cambridge, MA 02139, USA;
\supit{m}Lawrence Livermore National Laboratory, Livermore CA, 94550, USA;
\supit{n}Institute of Astronomy, Cambridge University, Cambridge,CB3 0HA, UK;
\supit{o}Department of Physics, Yale University,  New Haven, CT 06520, USA;
\supit{p}Department of Physics, University of Durham, Durham City, DH1 3LE, UK;
\supit{q}Noqsi Aerospace, Pine, CO 80470, USA;
\supit{r}RIKEN, Saitama 351-0198, Japan;
\supit{s}Department of Physics, Tokyo Metropolitan University, Tokyo 192-0397, Japan;
\supit{t}Harvard-Smithsonian Center for Astrophysics, Cambridge MA 02138, USA;
\supit{u}Faculty of Mathematics and Physics, Kanazawa University, Ishikawa  920-1192, Japan;
\supit{v}Department of Physical Science, Hiroshima University, Hiroshima 739-8526, Japan;
\supit{w}Physics Department, University of Miami, Coral Gables. FL 33124, USA;
\supit{x}Department of Astronomy and Physics, Saint Mary's University, Halifax, Nova Scotia, B3H 3C3, Canada;
\supit{y}Department of Applied Physics, University of Miyazaki, Miyazaki 889-2192, Japan;
\supit{z}Department of Physics, University of Tokyo, Tokyo 113-0033, Japan;
\supit{aa}Department of Physics and Astronomy, Rutgers University, Piscataway, NJ 08854, USA;
\supit{ab}Department of Physics and Astronomy, Johns Hopkins University, Baltimore, MD 21218, USA;
\supit{ac}Department of Physics, Chuo University, Tokyo 112-8551, Japan;
\supit{ad}Faculty of Human Development, Kobe University, Hyogo 657-8501, Japan;
\supit{ae}Department of Applied Quantum Physics and Nuclear Engnieering, Fukuoka 819-0395 Japan;
\supit{af}Research Institute for Science and Engineering, Waseda University, Tokyo 169-8555, Japan;
\supit{ag}Department of Physics, Tokyo Institute of Technology, Tokyo 152-8551, Japan;
\supit{ah}Gunma Astronomical Observatory, Gunma 377-0702, Japan;
\supit{ai}Department of Physics, Rikkyo University,Tokyo 171-8501, Japan;\\
\supit{aj}Department of Physics, Toho University, Chiba 274-8510, Japan;
\supit{ak}Department of Physics, Kougakuin University, Tokyo 192-0015, Japan;
\supit{al}School of Systems Engineering, Kochi University of Technology,  Kochi 782-8502, Japan;
\supit{am}Space Exploration Development Space Exploration, Canadian Space Agency, Saint-Hubert QC J3Y 8Y9, Canada;\\
\supit{an}Department of Physics and Department of Astronomy, Kyoto University, Kyoto 606-8502, Japan;
\supit{ao}Department of Electronic Information Systems, Shibaura Institute of Technology, Saitama 337-8570, Japan;
\supit{ap}IRFU/Service d'Astrophysique, CEA Saclay, France;\\
\supit{aq}Space Science Telescope Institute, Baltimore, MD 21218 USA;
\supit{ar}ESTEC, Noordwijk, The Netherlands;
\supit{as}Department of Physics, Tokyo University of Science, Tokyo 162-8601, Japan;
\supit{at}Department of Physics, University of Wisconsin, Madison, WI 53706, USA;\\
\supit{au}University of Waterloo, Waterloo, Ontario, N2L 3G1, Canada;
\supit{av}Department of Astronomy, University of Michigan, Ann Arbor, MI 48109,USA;
\supit{aw}Department of Mechanical Engineering, Chubu University, Aichi 487-8501, Japan;
\supit{ax}NASA/Marshall Space Flight Center, Huntsville, AL 35812, USA;
\supit{ay}JAXA, Ibaraki 305-8505, Japan;\\
\supit{az}Department of Physics, Nara Women's University, Nara 630-8506, Japan;\\
\supit{ba}Columbia Astrophysics Laboratory, Department of Astronomy, Columbia University, New York, NY 10027, USA;
\supit{bb}European Space Agency (ESA) European Space Astronomy Centre (ESAC), Madrid, Spain;
\supit{bc}Department of physics and Astronomy, University of Manitoba, Winnipeg, MB R3T 2N2, Canada;
\supit{bd}Department of Physics, Saitama University, Saitama 338-8570, Japan;
\supit{be}Faculty of Social and Information Sciences, Nihon Fukushi University, Aichi 475-0012, Japan;
}
\authorinfo{ASTRO-H Web site: http://ASTRO-H.isas.jaxa.jp/index.html.en}

\begin{document} 
  \maketitle 

\begin{abstract}

The joint JAXA/NASA ASTRO-H mission is the sixth in a 
series of highly successful X-ray missions initiated by 
the Institute of Space and Astronautical Science (ISAS). 
ASTRO-H will investigate the physics of the high-energy 
universe via a suite of four instruments, covering a very 
wide energy range, from 0.3~keV to 600~keV.  These instruments 
include a high-resolution, high-throughput spectrometer 
sensitive over 0.3--12~keV with high spectral resolution of 
$\Delta E $ $\leqq$ 7~eV, enabled by a micro-calorimeter 
array located in the focal plane of thin-foil X-ray optics;  
hard X-ray imaging spectrometers covering 5--80~keV, located in 
the focal plane of multilayer-coated, focusing hard X-ray mirrors;  
a wide-field imaging spectrometer sensitive over 0.4--12~keV, 
with an X-ray CCD camera in the focal plane of a soft X-ray 
telescope;  and a non-focusing  Compton-camera 
type soft gamma-ray detector, sensitive in the 40--600~keV band. The simultaneous 
broad bandpass, coupled with high spectral resolution, will enable 
the pursuit of a wide variety of important science themes.  

\end{abstract}




\section{Introduction}

ASTRO-H, which was formerly called NeXT, is an international X-ray satellite 
that Japan plans to launch with the H-II A rocket in 2014\cite{Ref:Proposal,Ref:Proposal03,Ref:Kunieda2004,Ref:Takahashi,Ref:Takahashi2008,Ref:Takahashi2010}.  
NASA has selected the US participation in  ASTRO-H as a Mission of Opportunity in the Exploration Program.   
Under this program, the NASA/Goddard Space Flight Center collaborates with 
ISAS/JAXA on the implementation of an X-ray micro-calorimeter and soft X-ray telescopes
(SXS Proposal NASA/GSFC, 2007)\cite{Ref:ProposalNASA}. Other international members are from Stanford 
University, SRON, Geneva University,CEA/DSM/IRFU, CSA and ESA. In early 
2009, NASA, ESA and JAXA have selected science advisors to provide 
scientific guidance to the ASTRO-H project relative to the design/development 
and operation phases of the mission. The ESA contribution to the ASTRO-H Mission includes the procurement of payload hardware elements 
which enhance the scientific capability of the mission.

The history and evolution of the Universe can be described as a process in which structures 
of different scales  such as stars, galaxies, and clusters of galaxies are formed.
In some cases, during this process, the matter and energy concentrate to an extreme degree in the form 
of black holes and neutron stars.  It is a mystery of Nature why and how  the
overwhelming diversity over orders of magnitude  in spatial and density scales has been produced in the Universe
following an expansion from a nearly uniform state. One of the best probes of this process are clusters of galaxies, the largest astronomical objects in
the Universe. Observing clusters of galaxies and revealing their history will lead to an understanding
of how the largest structures form and evolve in the Universe. Equally important is  studying how supermassive black holes form and develop, and what a role they play in forming  galaxies and clusters of galaxies.

\begin{figure}
\centerline{\includegraphics[scale=0.55,clip]{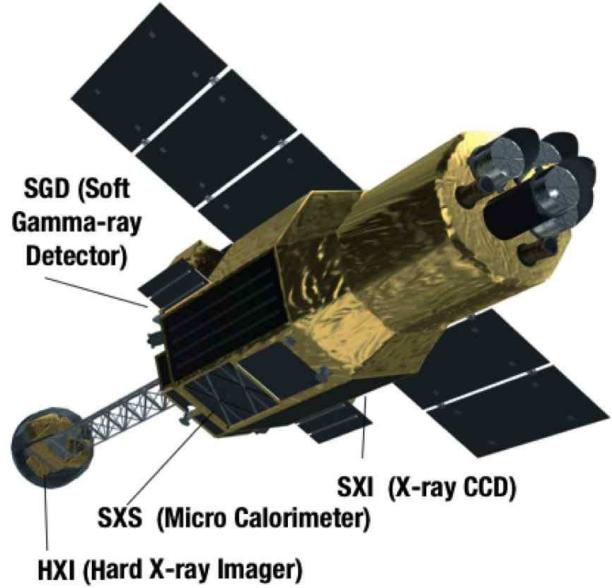}}
\caption{Schematic view of the \astroh\ satellite. The total mass at launch will be $\sim 2700$\,kg. \astroh\ will be launched into a circular orbit with altitude of $500-600$\,km, and inclination of $\sim 31$\,degrees.}
\label{Fig:ASTRO-H}
\end{figure}

\begin{table}
\caption{ASTRO-H Mission}
\begin{center}
\begin{tabular}{ll}
\hline
 Launch site & 
Tanegashima Space Center, Japan\\
 Launch vehicle& JAXA H-IIA rocket\\
 Orbit Altitude& $\sim$550 km\\
 Orbit Type& Approximate circular orbit\\
 Orbit Inclination& $\sim$ 31 degrees\\
 Orbit Period& 96 minutes\\
Total Length& 14 m\\
 Mass& $\sim$ 2.7  metric ton\\
 Power& $<$ 3500 W\\
 Telemetry Rate& 8 Mbps (X-band QPSK) \\
 Recording Capacity& 12 Gbits at EOL\\
 Mission life & $>$ 3 years\\
\hline
 \end{tabular}
 \end{center}
\end{table}
 The X-ray band is capable of probing extreme environments of the Universe such as those near black holes or the surface of neutron stars, as well as observing exclusively the emission from high-temperature gas and selectively the emission from accelerated electrons. In recent years, \chandra, \xmm, \suzaku\ and other X-ray missions have made great advances in X-ray Astronomy. We have obtained knowledge that has revolutionized our understanding of the high energy Universe and we have learned that phenomena observed in the X-ray band are deeply connected to those observed in other wavelengths from radio to $\gamma$-rays.

In order to revolutionize X-ray astronomy even further, the ASTRO-H mission is equipped with a suite of sensitive
instruments with the highest energy resolution ever achieved at E $>$ 3 keV and a wide energy range spanning
four decades in energy from soft X-rays to gamma-rays (Fig. \ref{Fig:ASTRO-H} ). This instrumental suite will provide the best sensitivity ever achieved for spectroscopy in the 1$-$600keV band. 
The mission aims to understand the dynamics of the Universe in
general, and study compact regions of high matter and energy
concentration allowing probe of production of energetic particles,
which are far from the thermal equilibrium.
%

\section{Science Requirements}

ASTRO-H is advancing the technologies developed through a series of
highly successful X-ray missions initiated in ISAS, beginning with
the launch of the Hakucho mission in 1979 through to the currently
operating Suzaku mission.
The prime scientific goal for ASTRO-H is to address a number of fundamental questions in contemporary astrophysics, as listed below.
\begin{center}
\bf{Scientific Goals and Objectives}
\end{center}

\subsubsection*{Revealing the large-scale structure of the Universe and its evolution}

\begin{itemize}
\item 
ASTRO-H will observe clusters of galaxies, the largest bound structures in the Universe, with the aim
to reveal the interplay between the thermal energy of the intracluster medium and the kinetic energy
of sub-clusters, from which clusters form; measure the non-thermal energy and chemical composition; and to directly trace the
dynamic evolution of clusters of galaxies.

\item ASTRO-H will observe distant supermassive black holes hidden 
by thick intervening material  with  100 times higher sensitivity 
than the currently operating Suzaku, and will study their evolution and a role they play in galaxy formation.
\end{itemize}

\subsubsection*{Understanding the extreme conditions in the Universe}

\begin{itemize}
\item ASTRO-H will measure the motion of matter very close to black 
holes with the aim to sense the gravitational distortion of  space  to understand 
the structure of relativistic space-time and to study the physics of the accretion process.
\end{itemize}

\subsubsection*{Exploring the diverse phenomena of the non-thermal Universe}
\begin{itemize}
\item ASTRO-H will derive the physical conditions of the sites where 
high energy cosmic ray particles gain energy and will elucidate 
the processes by which gravity, collisions, and stellar explosions energize  
those cosmic rays.

\end{itemize}

\subsubsection*{Elucidating dark matter and dark energy}
\begin{itemize}
\item ASTRO-H will map the distribution of dark matter in clusters 
of galaxies and will determine the total mass of galaxy clusters 
at different distances (and thus at different ages), and will study the role 
of dark matter and dark energy in the evolution of these systems.

\end{itemize}

In order to achieve the cutting-edge scientific goals
described above, ASTRO-H is designed with the most advanced
technologies.
With an unprecedented spectroscopic capability and a wide-band energy coverage, ASTRO-H
will measure the motion of hot gas, depicting the dynamic nature of the 
evolution of the Universe.  These measurements will be the  key in the great pursuit 
to understand the origin of the dark matter filling  the Universe.

\section{Spacecraft and Instruments}

\begin{figure}
\centerline{\includegraphics[scale=0.35,clip]{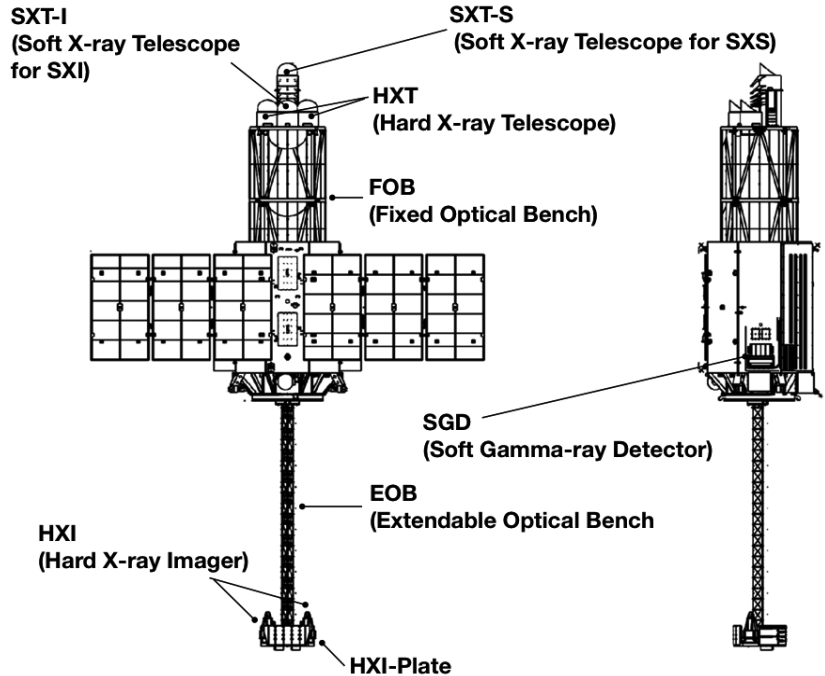}}
\caption{Schematic view of the \astroh\ satellite. }
\label{Fig:ASTRO-H1}
\end{figure}

 The ASTRO-H X-ray observatory will consist of four focusing telescopes mounted on a fixed optical bench (FOB).  Two of the four telescopes are Soft X-ray Telescopes (SXTs) and they have a 5.6 m focal length (Fig. \ref{Fig:ASTRO-H1}).  They will focus medium-energy X-rays (E $\sim$ 0.3-12 keV) onto focal plane detectors mounted on the base plate of the FOB.  One SXT will point to a micro-calorimeter spectrometer array with
excellent energy resolution of $\leqq$7 eV, and the other SXT will point to
a large-area CCD array. 
The other two telescopes are Hard X-ray Telescopes (HXTs) capable of focusing high-energy X-rays (E = 5-80 keV).  The focal length of the HXTs is 12 m.   The Hard X-ray Imaging detectors (HXIs) are mounted at the end of a 6 m extendable optical bench (EOB) that is stowed to fit in the launch fairing and deployed once in orbit.  
 In order to extend the energy 
coverage to the soft $\gamma$-ray region up to 600 keV, the Soft Gamma-ray 
Detector (SGD) will be implemented as a non-focusing detector.  
  Two SGD detectors, each consisting of three units
 will be mounted separately on two sides of the satellite. With these instruments, 
ASTRO-H will cover the entire bandpass between 0.3~keV and 600~keV.  The key parameters  of 
those instruments based on the base-line design are 
summarized in Table.~\ref{Table:Spec}.

The lightweight design of the EOB renders it to be vulnerable to distortions from thermal fluctuations in low-Earth orbit (LEO) and spacecraft attitude manoeuvres.  Over the long exposures associated with X-ray observing, such fluctuations might impair HXI image quality unless a compensation technique is employed. To provide the required corrections, the Canadian contribution to the the ASTRO-H project is  a laser metrology system (the Canadian ASTRO-H Metrology System, CAMS) that will measure
displacement in the alignment of the HXT optical path. The CAMS consist of 
 a laser and detector module (CAMS-LD) located on the top plate of the FOB, and a passive target module (CAMS-T) consisting of a retroreflector (corner cube mirror) mounted on the EOB detector plate\cite{Ref:Gallo}.

\begin{table}
\caption{Key parameters of the ASTRO-H payload}
\begin{center}
\label{Table:Spec}
\begin{footnotesize}
\begin{tabular}{|l|p{2.5cm}|p{2.5cm}|p{2.4cm}|p{2.8cm}|}
\hline
Parameter & Hard X-ray  & Soft X-ray & Soft X-ray   & Soft $\gamma$-ray  \\
&  Imager & Spectrometer  &  Imager & Detector\\
&   (HXI) & (SXS)  &   (SXI) & (SGD)\\
\hline
\hline
Detector & Si/CdTe & micro& X-ray & Si/CdTe  \\
 technology & cross-strips & calorimeter & CCD & Compton Camera \\
 \hline
Focal length & 12 m & 5.6 m & 5.6 m  & -- \\
\hline
Effective area & 300 cm$^{2}$@30 keV  &  210 cm$^{2}$@6 keV  &  360 cm$^{2}$@6 keV & 
 $>$20~cm$^{2}$@100 keV  \\
 &  & 160 cm$^{2}$ @ 1 keV  &    &  Compton Mode  \\
\hline
Energy range & 5 --80 keV & 0.3 -- 12 keV  & 0.5 -- 12 keV & 40 -- 600 keV \\
\hline
Energy  & 2 keV& $<$ 7 eV & $<$ 200 eV& $<$ 4 keV \\
resolution& (@60 keV)  & (@6 keV) & (@6 keV) &  (@60 keV) \\
 (FWHM) & &  & & \\
 \hline
Angular  & $<$1.7 arcmin& $<$1.3 arcmin& $<$1.3 arcmin& -- \\
resolution & & & & \\
\hline
Effective & $\sim$ 9 $\times$ 9  & $\sim$ 3 $\times$ 3   & $\sim$ 38 $\times$ 38 & 0.6 $\times$ 0.6 deg$^{2}$ \\
 Field of View &arcmin$^{2}$ & arcmin$^{2}$& arcmin$^{2}$& ($<$ 150 keV) \\
\hline
Time resolution  & 25.6  $\mu$s & 5  $\mu$s &  4 sec/0.1 sec & 25.6  $\mu$s \\
\hline
Operating  & $-$20$^{\circ}$C & 50 mK & $-$120$^{\circ}$C& $-$20$^{\circ}$C\\
 temperature  & & & & \\
 \hline
\end{tabular}
\end{footnotesize}
\end{center}
\end{table} 

In the following sections, these instruments  are briefly described.
Detailed descriptions of the instruments and their current status are available  in other papers in these  proceedings
\cite{Ref:Awaki2012,Ref:Fujimoto2012,Ref:Okajima2012,Ref:Kokubun2012,Ref:Tsunemi2012,Ref:Watanabe2012,Ref:FW2012}. 

\subsection{Soft X-ray Spectrometer System}

\begin{figure}
\centerline{\includegraphics[height=7cm,clip]{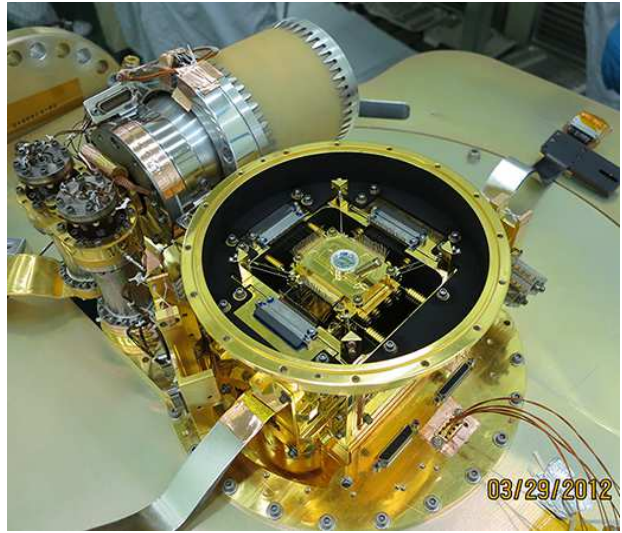} }
\caption{The SXS engineering model detector assembly.  At the center of the assembly is the x-ray calorimeter housing.  This is suspended from the outer structure using Kevlar, and electrical connections to the housing are made using tensioned wires to reduce the sensitivity to microphonics.  At the center of the calorimeter housing is an aluminum/polyimide blocking filter and a $^{55}$Fe calibration source used to illuminate a dedicated calibration pixel for monitoring the absolute gain.  The overall assembly is about 12.7 cm in diameter.}
\label{Fig:XRS_EM1}
\end{figure}
\begin{figure}
\centerline{\includegraphics[height=6cm,clip]{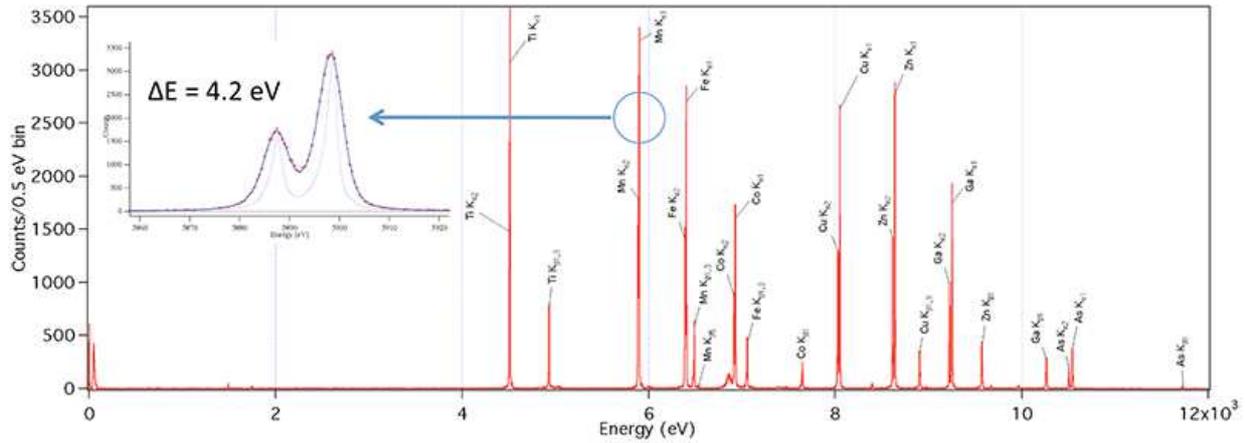}  }
\caption{Laboratory X-ray spectrum obtained with the Astro-H Soft X-Ray Spectrometer engineering model detector assembly.  The spectrum shows the enormous spectral dynamic range that can be obtained.  The spectral resolution is 4.2 eV over the entire array, and is achieved over the full energy range where astrophysically abundant atomic transitions will be detected (less than about 8 keV), providing a resolving power of about about 1400 at 6 keV.  The required resolution is 7 eV.}
\label{Fig:XRS_EM2}
\end{figure}

The soft X-ray Spectrometer (SXS) consists of the Soft X-ray Telescope 
(SXT), the X-ray Calorimeter Spectrometer (XCS) and the cooling system\cite{Ref:Mitsuda,Ref:Mitsuda2010}.
The XCS is a 36-pixel system with an energy resolution of $\leq$7~eV between 0.3--12~keV(Fig. \ref{Fig:XRS_EM1}). 
The array design for the SXS is basically the same as that for the Suzaku/XRS, but has larger pixel pitch and absorber size.  HgTe absorbers are attached to ion-implanted Si thermistors formed on suspended Si micro-beams.  The array has 814 $\mu$m pixels on an 832 $\mu$m pitch and was manufactured during the Suzaku/XRS program along with arrays with smaller pixel size as an option for a larger field of view.  
For ASTRO-H, the longer focal length of the SXS (5.6 m vs. 4.5 m for Suzaku) necessitated the use of these larger arrays to maintain a FOV of at least 2.9 $\times$ 2.9 arcminutes.  The 8.5-micron-thick absorbers were fabricated by EPIR Corporation and diced using reactive ion etching.  These  absorbers provide high quantum efficiency across the 
0.3--12~keV band. Despite the larger pixel size for the SXS (factor of 1.7 in volume), the energy resolution is substantially improved from $\sim$ 6 eV to $\sim$ 4 eV (FWHM) (Fig. \ref{Fig:XRS_EM2}).  The main reasons for this are that EPIR developed a process to produce HgTe with lower specific heat and that the operating temperature of the instrument has been lowered from 60 mK to 50 mK\cite{Ref:Porter2}.

With a 5.6-m focal length, the 0.83~mm pixel pitch corresponds to 
0.51~arcmin, giving the array a field of view of 3.05~arcmin on a side.  
The detector assembly provides electrical, thermal, and mechanical interfaces 
between the detectors (calorimeter array and anti-coincidence particle detector) 
and the rest of the instrument.  The SXT for the XCS
is an upgraded version of the Suzaku X-ray telescope with  improved
angular resolution and  larger collecting area\cite{Ref:Okajima,Ref:Selemistos2010}.
The SXS effective 
area at 6~keV will be at least 210~cm$^{2}$, a 60~\% increase over the Suzaku XRS,  while at 1~keV 
the SXS has 160~cm$^{2}$, a 20~\% increase.

\begin{figure}
\centerline{{\includegraphics[height=8cm,clip]{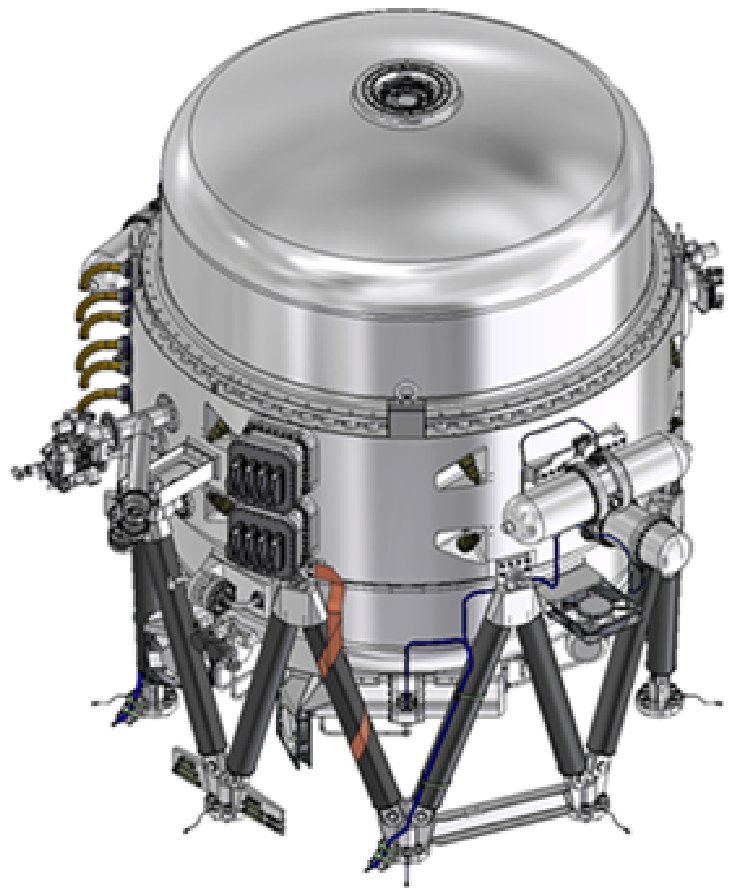}}{\includegraphics[height=8cm,clip]{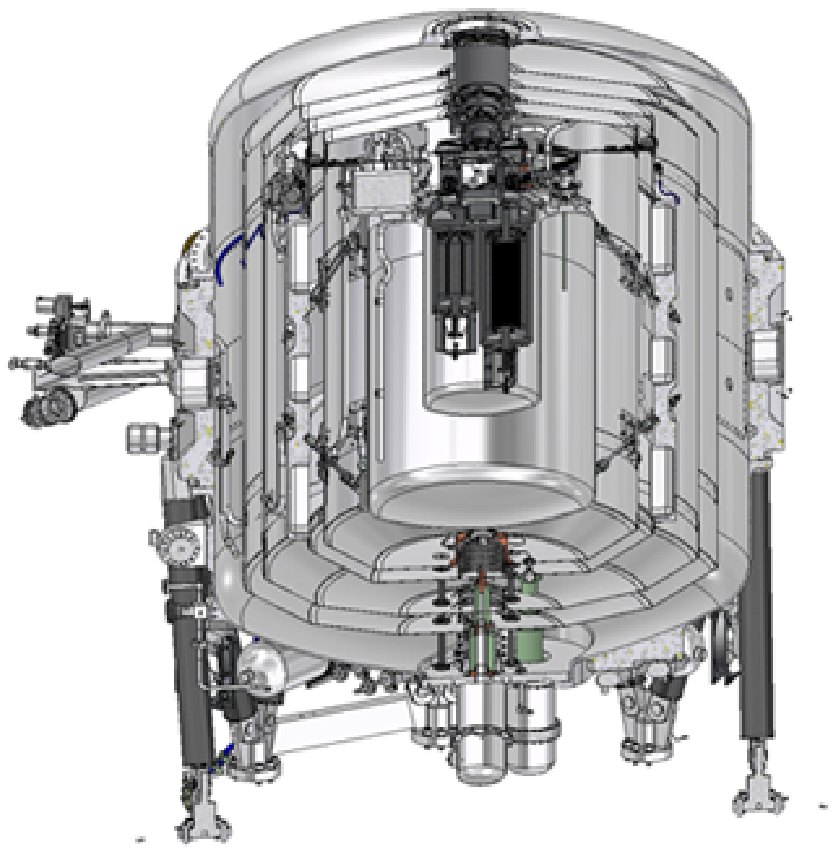}}}
\caption{Outlook   and cross sectional view of the SXS dewar. The outer shell of the dewar is 950 mm in diameter. }
\label{Fig:XRS}
\end{figure}

The XCS cooling system must cool the array to 50~mK with sufficient duty cycle to 
fulfill the SXS scientific objectives:  this requires extremely low heat loads.  
To achieve the necessary gain stability and energy resolution, the cooling 
system must regulate the detector temperature to within 2~$\mu$K rms for at 
least 24~hours per cycle\cite{Ref:Fujimoto2010,Ref:Fujimoto2012}.  From the detector stage to room temperature, the 
cooling chain is composed of a 3-stage  Adiabatic Demagnetization 
Refrigerator (ADR), superfluid liquid $^{4}$He (hereafter LHe), 
a $^{4}$He Joule-Thomson (JT) cryocooler, and 2-stage Stirling cryocoolers (Fig. \ref{Fig:XRS}.)  
An ADR has been adopted because it readily meets 
the requirements for detector temperature, stability, recycle time, 
reliability in the space environment, and previous flight heritage\cite{Ref:Shirron}.  The 
design of Stirling cryocoolers is based on coolers developed for space-flight 
missions in Japan (Suzaku, AKARI, and the SMILES instrument  deployed on 
the ISS\cite{Ref:Narasaki}) that have achieved an excellent performance with 
respect to cooling power, efficiency, long life and mass. 30 L of LHe is used
as a heat-sink for the 2-stage 
ADR. To reduce the parasitic heat load on the He tank, a 
$^{4}$He JT cryocooler is used to cool a 4 K shield. To achieve redundancy for failure 
(unexpected loss) of LHe, another  ADR (3rd stage ADR) is used between the 
He tank and the JT cryocooler, with two heat-switches on both sides.  
This ADR is operated if LHe is lost, to cool down the 1 K shield (He tank). 
A series of five blocking filters shield the calorimeter array from UV and 
longer wavelength radiation. The aluminized polyimide filters are similar 
to those successfully used on Suzaku. 

\begin{figure}
\centerline{\includegraphics[height=5.5cm,clip]{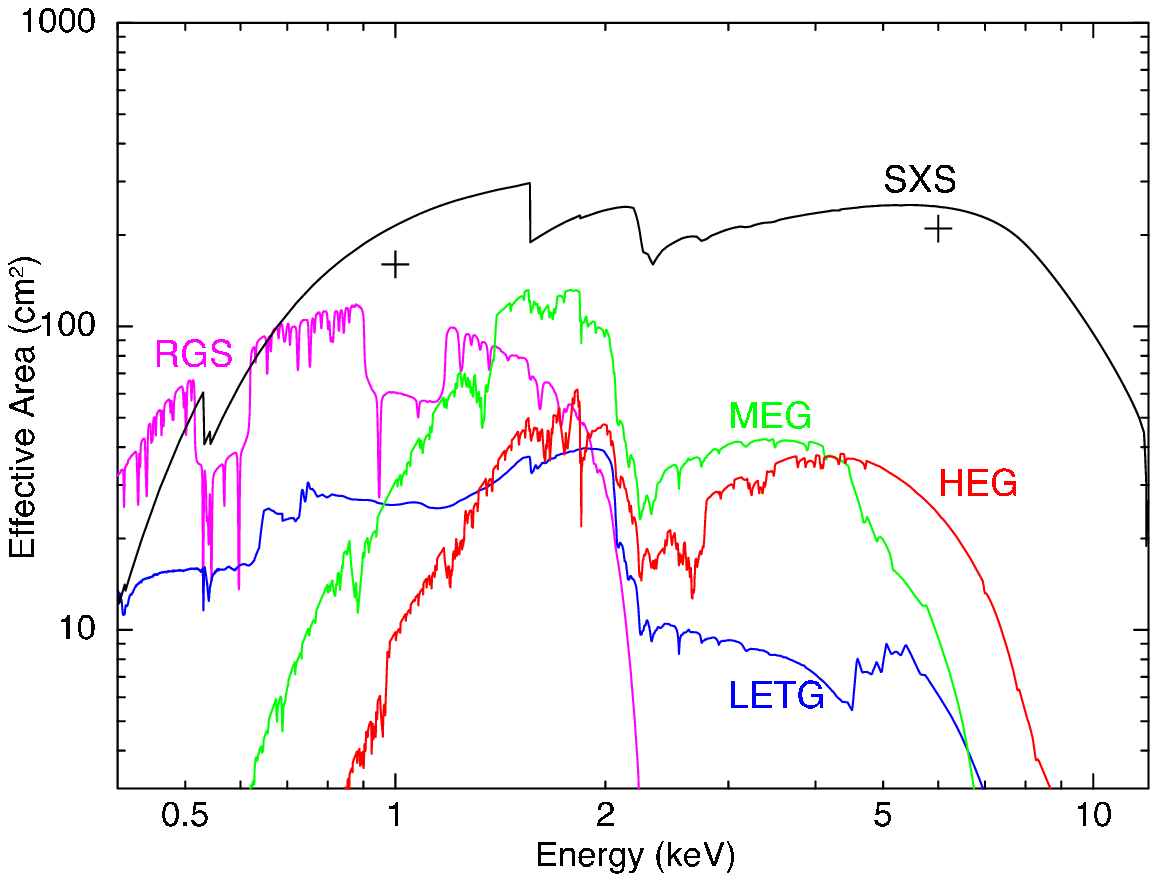}  \hspace{5mm} \includegraphics[height=5.5cm,clip]{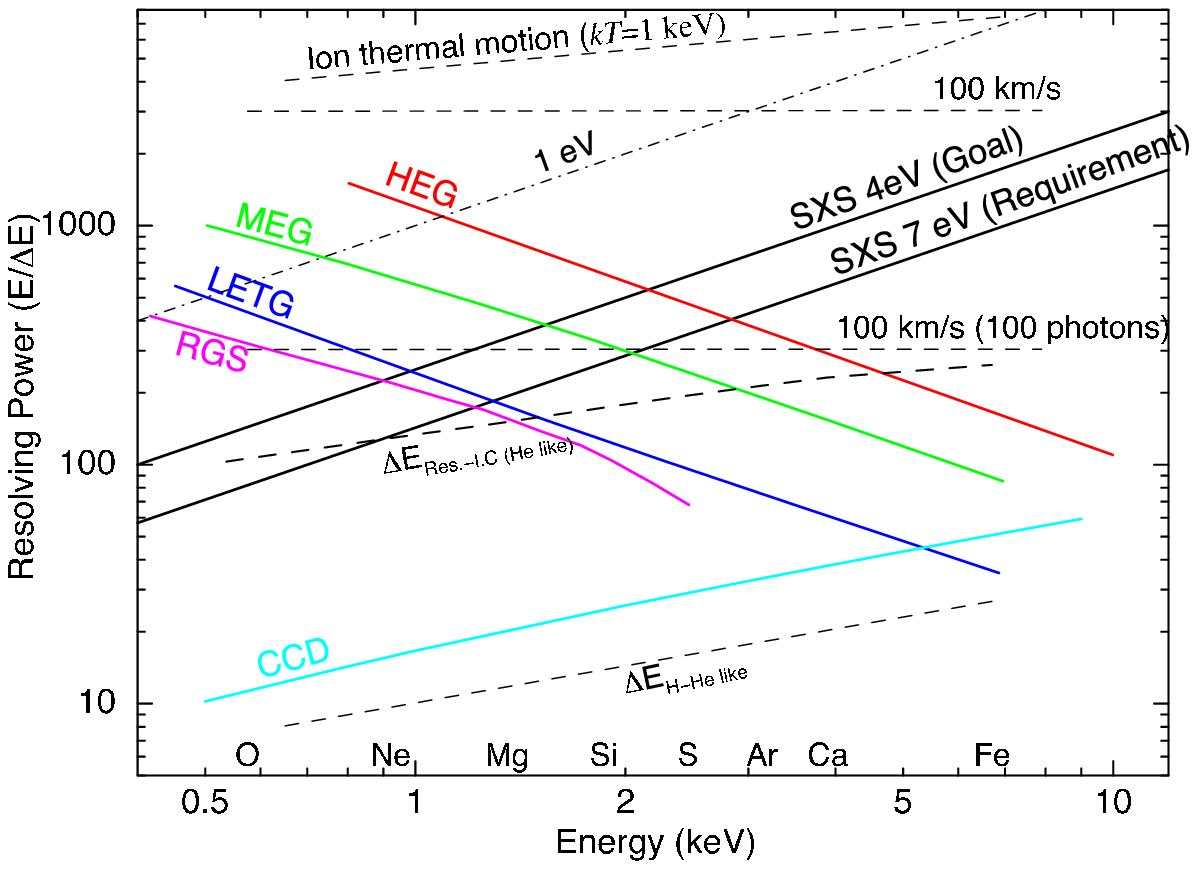}}
\caption{{\bf (a)} Effective areas of high-resolution X-ray spectroscopy missions as functions of X-ray energy. The curve for the \astroh\ SXS is the present best estimate for a point source. The two crosses show the mission requirements at specific energies. The \xmm\ RGS effective area is the summation of first order spectra of the two instruments (RGS-1 and RGS-2). The effective areas of the LETG, MEG and HEG onboard \chandra\ are summations of the first order dispersions in $\pm$ directions.
 {\bf (b)} Resolving power of the \astroh\ SXS as a function of X-ray energy for the two cases, 4\,eV resolution (goal) and 7\,eV (requirement). The resolving power of high resolution instruments onboard \chandra\ and \xmm\ and typical resolving power of X-ray CCD cameras are also shown for comparison \cite{Ref:Mitsuda2010}. }
\label{Fig:XRS2}
\end{figure}

In combination with a high throughput X-ray telescope, the 
SXS improves on the Chandra and XMM-Newton grating spectrometers in 
two important ways. At E $>$ 2~keV, SXS is both more sensitive and has 
higher resolution (Fig.\ref{Fig:XRS2}), especially in the Fe K band where 
SXS has 10 times larger the collecting area and much better energy resolution, giving a net 
improvement in sensitivity by a factor of 30 over Chandra.  The broad bandpass of 
the SXS encompasses the critical inner-shell emission and absorption lines 
of Fe I-XXVI between 6.4 and 9.1~keV. Fe K lines provide particularly useful 
diagnostics because of their (1) strength, due to the high abundance and large 
fluorescent yield (30\%), (2) spectral isolation from other lines, and (3) 
relative simplicity of the atomic physics. Fe K emission lines reveal conditions 
in plasmas with temperatures between 10$^{7}$ and 10$^{8}$ K, which are typical values for 
stellar accretion disks, SNRs, clusters of galaxies, and many stellar coronae.  
In cooler plasmas, Si, S, and Fe fluorescence and recombination occurs 
when an X-ray source illuminates nearby neutral material. Fe emission 
lines provide powerful diagnostics of non-equilibrium ionization 
due to inner shell K-shell transitions from Fe XVII--XXIV\cite{Ref:Decaux}.

The SXS uniquely performs high-resolution spectroscopy of extended sources. 
In contrast to a grating, the spectral resolution of the calorimeter is 
unaffected by source's angular size because it is non-dispersive.  
For all sources with angular extent larger than 30~arcsec, Chandra MEG 
energy resolution is degraded compared with that of a CCD; the energy 
resolution of the XMM-Newton RGS is similarly degraded for sources 
with angular extent $\geq$25 arcsec. SXS therefore makes possible high-resolution 
spectroscopy of sources inaccessible to current grating instruments.

In order to obtain a good performance for bright sources, 
a filter wheel (FW) assembly, which includes a wheel with selectable filters and a set of modulated X-ray sources,
 are provided by SRON and Univ. of Geneva.
It will be used at a distance of 90 cm from the detector. The FW is able to rotate a suitable filter into the
beam to optimize the quality of the data, depending on the source characteristics\cite{Ref:Vries,Ref:FW2012}. In addition to the
filters, a set of on-off-switchable X-ray calibration sources, using light sensitive photo-cathode, will be implemented.
With these calibration sources, it is possible to calibrate the energy scale with a typical 1$-$2 eV accuracy, and 
 will allow proper gain and linearity calibration of the detector in flight.

\subsection{Soft X-ray Imaging System}
\begin{figure}
\centerline{\includegraphics[height=8cm,clip]{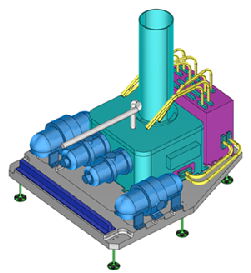}
\includegraphics[height=5cm,clip]{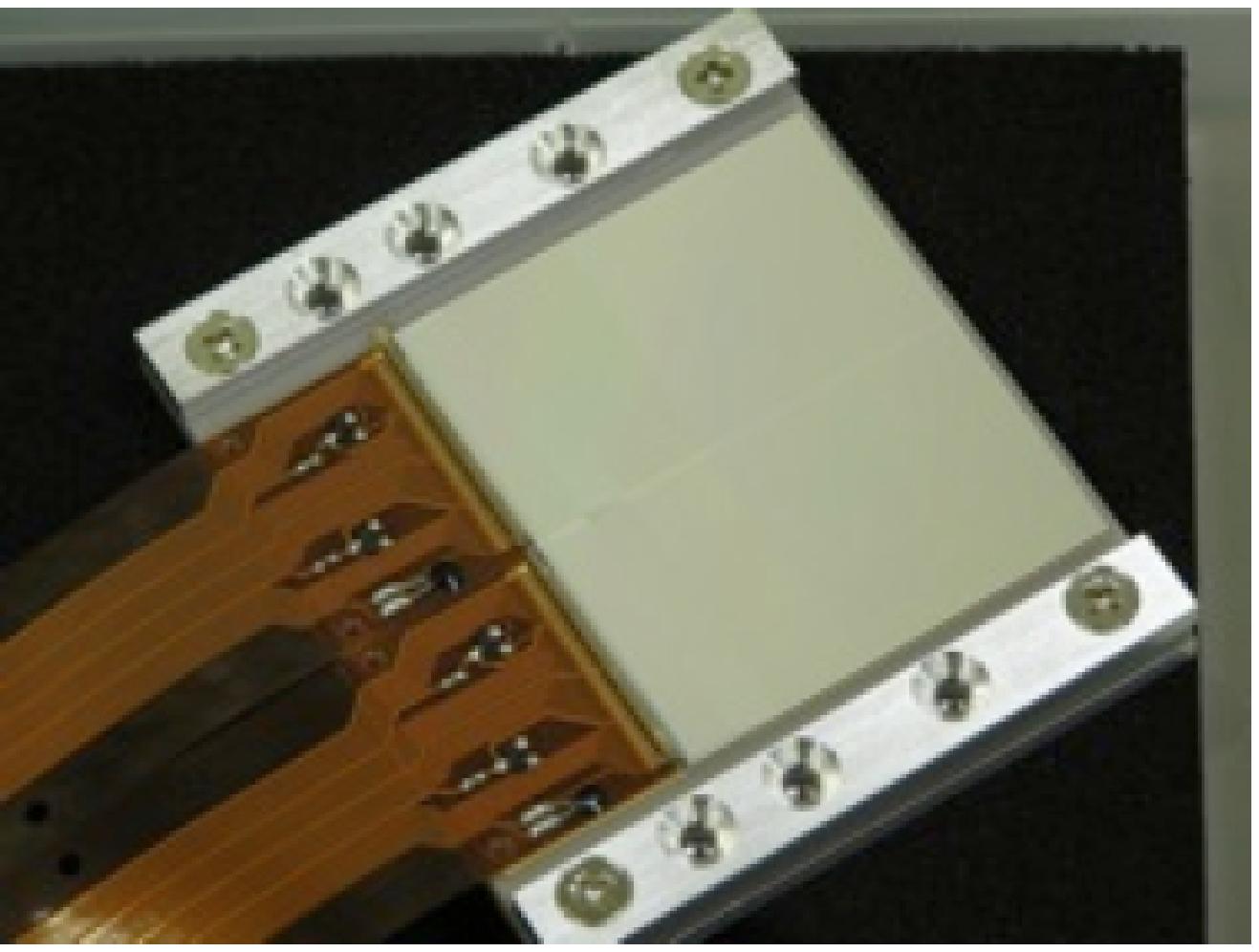}}
\caption{Schematic drawing of the Soft X-ray Imager and a picture of a prototype CCD chip. }
\label{SXI-OUTLOOK}
\end{figure}

X-ray sensitive silicon charge-coupled devices (CCDs) are   key detectors
for X-ray astronomy. The low background and high energy resolution
achieved with the XIS/Suzaku clearly show that the X-ray CCD can also
play a very important role in the ASTRO-H mission. The  soft X-ray imaging system 
will consist of an imaging mirror and a CCD camera, (Soft X-ray Telescope (SXT-I),  Soft X-ray Imager (SXI)) and the cooling system\cite{Ref:Tsuru,Ref:Tsunemi,Ref:Tsunemi2010,Ref:Tsunemi2012}. Fig.~\ref{SXI-OUTLOOK} 
 shows  a schematic  drawing of the
SXI.
 
\begin{figure}[htb]
\centerline{\includegraphics[height=5cm]{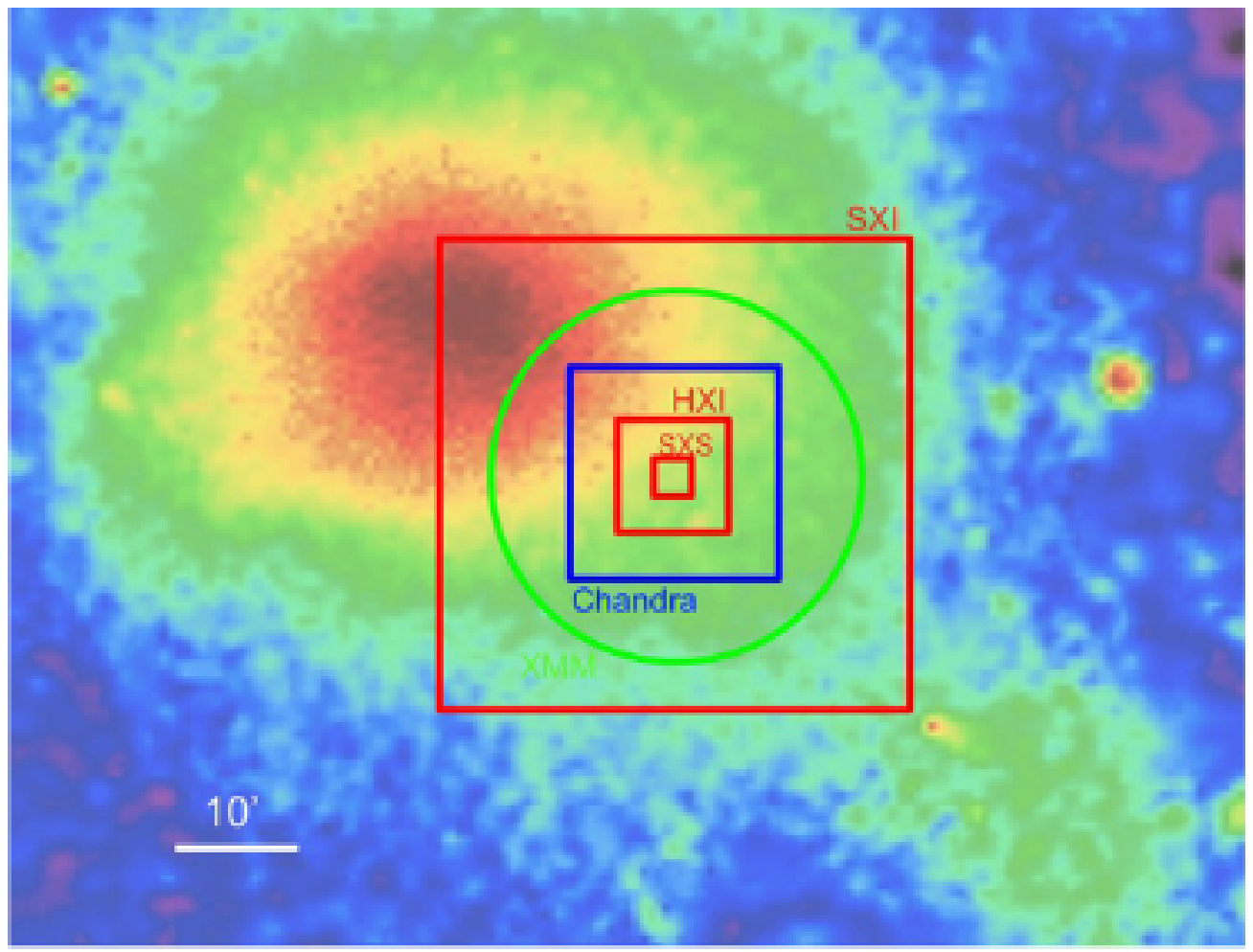}
\includegraphics[height=5cm]{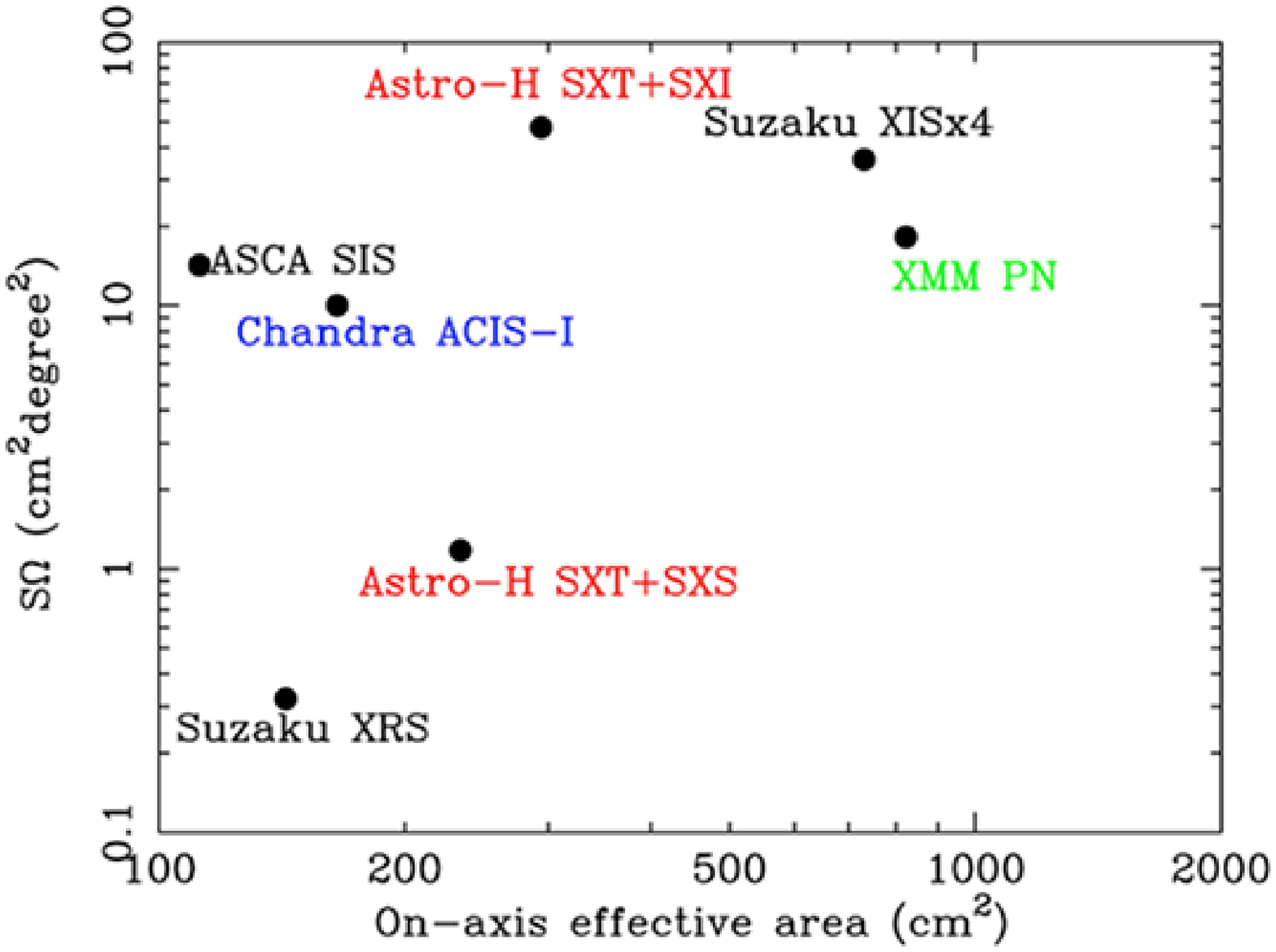}}
\caption{(left) Fields of view of the ASTRO-H instruments, SXS, SXI, HXI (the red boxes). Chandra ACIS-I and XMM are also shown for comparison. The background image is the Coma cluster taken with ROSAT (credit: ROSAT/MPE/S. L. Snowden). (right) Grasp vs on-axis effective area at 7 keV of SXT-I+SXI, SXT-S+SXS.  Suzaku XIS, Chandra ACIS-I, XMM PN are also shown for comparison.
}
\label{Fig:FOV}
\end{figure}

In order to cover the soft X-ray band below 12~keV,
the SXI will use next generation Hamamatsu CCD chips with
a thick depletion layer of 200 $\mu$m, low noise, and almost no cosmetic defects. 
The SXI features a large FOV and covers a 38$\times$38~arcmin$^{2}$ region on the sky,
complementing the smaller FOV of the 
SXS calorimeter (Fig. \ref{Fig:FOV}). A mechanical cooler ensures a long operational life at 
$-$120~$^\circ$C\@. The  quantum efficiency 
is better than the one achieved in Suzaku XIS over the entire 0.3$-$12 kev bandpass\@.  The imaging mirror has a 5.6-m focal 
length, and a diameter of 45~cm.
The nominal aim point will be placed at the center of SXS field of view, which is 4.3 arcmin offset from the SXI center. 

 \subsection{Hard X-ray Imaging System}

\begin{figure}[htb]
\centerline{\includegraphics[width=12cm]{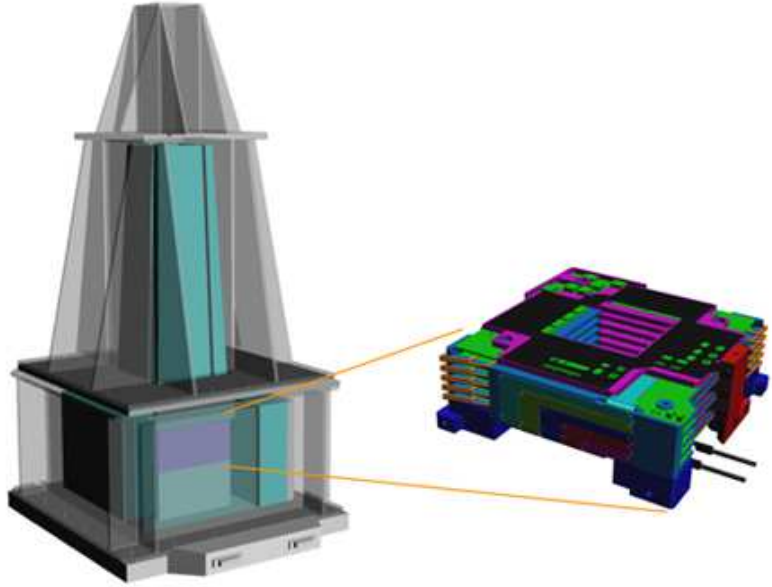}}
\caption{The Hard X-ray Imager. A stack of Si and CdTe double sided cross-strip detectors is
mounted in a well-type BGO shield.
}
\label{Fig:HXI}
\end{figure}

The hard X-ray imaging system onboard ASTRO-H consists of two identical mirror-detector 
pairs (HXT and  HXI).
The HXT has conical-foil mirrors with depth-graded multilayer reflecting surfaces 
that provide efficient reflection over 5--80~keV energy range\cite{Ref:Ogasaka,Ref:Kunieda2010,Ref:Awaki2012}.   The effective area of the HXT
is maximized for a long focal length. 

A depth-graded multilayer mirror reflects X-rays not only by total
external reflection but also by the Bragg reflection. In order to obtain
a high reflectivity up to 80~keV, the HXTs consist of a stack of
multilayers with different sets of periodic length and number of layer
pairs with a platinum/carbon coating. The technology of a hard X-ray
focusing mirror has already been proven by the balloon programs
InFOC$\mu$S (2001, 2004)\cite{Ref:Kunieda2006,Ref:Ogasaka:In}, 
HEFT (2004)\cite{Ref:Fiona} and SUMIT (2006)\cite{Ref:Kunieda2006}
and very recently with the NuSTAR satellite\cite{Ref:SPIE_NuSTAR}.

Recent test  of the first flight mirror  have validated the 
HXT design, achieving a collecting area of 174 cm$^2$ at 30 keV  for one telescope, with 
a focal length of 12 m.

The non-imaging instruments flown so far were essentially limited to
studies of sources with 10--100~keV fluxes of  $>$4 $\times$ 10$^{-12}$--%
10$^{-11}$~erg~cm$^{-2}$s$^{-1}$, at best. This limitation is due to the presence
of high un-rejected backgrounds from particle events and Cosmic X-ray
radiation, which increasingly dominate above 10~keV. Imaging, and
especially focusing instruments have two tremendous advantages. Firstly,
the volume of the focal plane detector can be made much smaller than for
non-focusing instruments, thus reducing the absolute background level
since the background flux generally scales with the size of the
detector.  Secondly, the residual background, often time-variable, can
be measured simultaneously with the source, and can be reliably
subtracted.  For these reasons, a focusing hard X-ray telescope in
conjunction with an imaging detector sensitive for hard
X-ray photons is the appropriate choice to achieve a breakthrough in
sensitivity for the field of high energy astronomy. In addition to the improvement
of sensitivity, the HXI provides a true imaging capability which enable us to
study spatial distributions of hard X-ray emission.

The HXI sensor consists of four layers of 0.5~mm thick Double-sided Silicon Strip
Detectors (DSSD) and one layer of 0.75~mm thick CdTe imaging 
detector (Fig.~\ref{Fig:HXI})\cite{Ref:Takahashi_SPIE1,Ref:Nakazawa,Ref:Kokubun,Ref:Kokubun2010,Ref:Kokubun2012}.  
In this configuration, soft X-ray photons below  $\sim$ 20 keV are 
absorbed in the Si part (DSSD), while hard X-ray photons above  $\sim$ 20 keV go through the
Si part and are detected by a newly developed CdTe double sided cross-strip detector. 
The low energy spectrum, obtained with Si, is less contaminated by 
the background due to activation in heavy material, such as Cd and Te.  
The DSSDs cover the energy below 30~keV while the 
CdTe strip detector covers the 20--80~keV band.
Each DSSD has a size of 3.2$\times$3.2~cm$^{2}$
and a thickness of 0.5~mm, resulting in 2~mm in total thickness, the same as that of the PIN detector 
of the HXD onboard Suzaku. A CdTe strip
detector has a size of 3.2$\times$3.2~cm$^{2}$ and a thickness of 0.75~mm. 
In addition to the increase of efficiency, the stack configuration and
individual readout provide information on the interaction depth. This
depth information is very useful to reduce the background in space
applications, because we can expect that low energy X-rays interact in
the upper layers and, therefore, it is possible to reject the low energy events
detected in lower layers. Moreover, since the background rate scales
with the detector volume, low energy events collected from the first few
layers in the stacked detector have a high signal to background ratio,
in comparison with events obtained from a monolithic detector with a
thickness equal to the sum of all layers.  

In the energy band above 10 keV, the number of photons from the source decreases and the detector background becomes the major limitations of its sensitivity. Since a significant fraction of the background events originate from interactions of the cosmic-ray with the detector structure, a tight active shield to reject cosmic-ray induced events is critical. 
Fast timing
response of the silicon strip detector and CdTe strip detector allows us to
place the entire detector inside a very deep well of an active shield
made of BGO (Bi$_{4}$Ge$_{3}$O$_{12}$) scintillators. The signal from the BGO
shield is used to reject background events. 
The BGO scintillator crystal, with its high density and high absorbing power, is commonly used as a material for anti coincidence shields in recent hard X-ray/soft gamma-ray instruments, such as INTEGRAL and Suzaku/HXD. In ASTRO- H/HXI and SGD, a 3-4 cm thick BGO crystal is employed. With this thickness, it can not only tag the cosmic-rays, but also reduce the number of protons hitting the detector material (e.g., CdTe) by its passive shielding and hence it reduces the background.

\subsection{Soft Gamma-ray Detector (SGD)}

Highly sensitive observations in the energy range above the HXT/HXI bandpass are crucial to study the
spectrum of X-rays arising from accelerated particles and black holes. 
In order to extend the energy coverage to the soft $\gamma$-ray region up to 600\,keV, the Soft Gamma-ray Detector (SGD) will be implemented as a non-focusing detector onboard \astroh \cite{Ref:Tajima2010,Ref:Watanabe2012}. The SGD measures soft $\gamma$-rays via reconstruction of the Compton scattering in the Compton camera, covering  an energy range of $40-600$\,keV with sensitivity at $300$\,keV of more than 10 times that of the \suzaku\ Hard X-ray Detector. It outperforms previous soft-$\gamma$-ray instruments in background
rejection capability by adopting a new concept of narrow-FOV Compton
telescope\cite{Ref:Takahashi_Yokohama,Ref:Takahashi_SPIE2}.

In order to lower the background dramatically and thus to improve the
sensitivity as compared to the HXD of Suzaku, the design combines a stack of Si
and CdTe pixel detectors to form a Compton camera (Si/CdTe Compton Camera)\cite{Ref:Takahashi_SiCdTe}.  The
telescope is then mounted inside the bottom of a well-type active
shield. Above $\sim$40~keV, each valid event is required to interact twice in the stacked
detector, once by Compton scattering in a stack of Si strip detectors,
and then by photo-absorption in the CdTe part (Compton mode).  Once the
locations and energies of the two interactions are measured, the Compton
kinematics allows the calculation of the energy and direction (as a cone in
the sky) of the incident $\gamma$-ray.    The major advantage of employing a narrow FOV is that
the direction of incident $\gamma$-rays is constrained to be inside the FOV. If the Compton cone
domes not intercept the FOV, we can reject the event as background. Most of the background can be
rejected by requiring this condition\cite{Ref:Takahashi_Yokohama}.

\begin{figure}
\centerline{\includegraphics[height=7.0cm,clip]{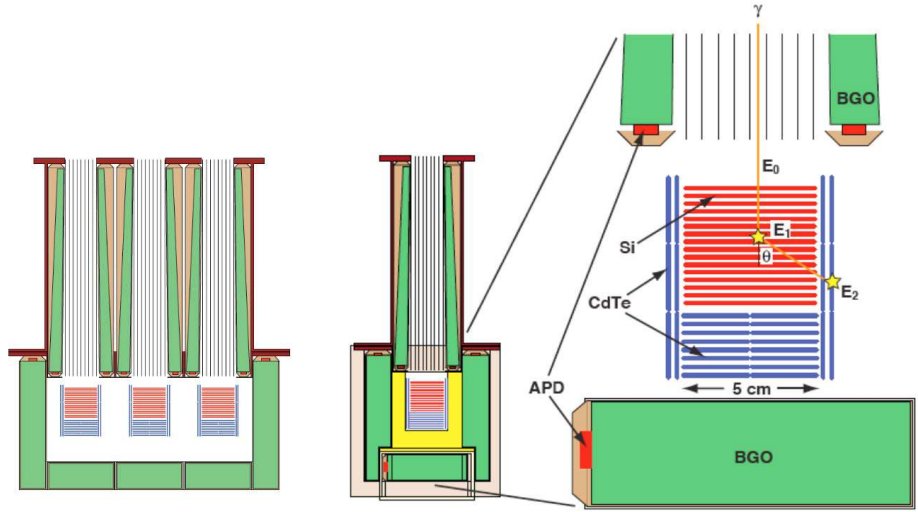}}
\caption{Conceptual drawing of an SGD Compton camera unit (green sections are the BGO anti- coincidence shields, red planes are the Si strip detectors in which the Compton scattering, occurs and the blue parts are the CdTe section in which the photons are absorbed.}
\label{Fig:SGD_Concept}
\end{figure}

As shown schematically in Fig.~\ref{Fig:SGD_Concept}, the detector
consists of 32 layers of 0.6~mm thick Si pad detectors 
and eight layers of CdTe pixellated detectors with a thickness of 0.75~mm.  The sides are 
also surrounded by two layers of CdTe pixel detectors. The 
opening angle provided 
by the BGO shield is $\sim$10~degrees at 500~keV. 
As 
compared to the HXD, the shield part is made compact by adopting the 
newly developed avalanche photo-diode.  
An additional PCuSn collimator restricts the field of view of the
telescope to 30' for photons below 100 keV  to minimize the
flux due to the Cosmic X-ray Background in  the FOV.  These modules
are then arrayed to provide the required area. 

The effective area of the SGD is $>$20~cm$^{2}$ at 100~keV in the
Compton mode.
It should be noted that when the Compton condition is not used,
the stacked DSSD  can be used as a standard photo-absorption type 
detector with  the total thickness $\sim$20~mm of silicon. The detector then
covers energies above 10~keV as a collimated-type $\gamma$-ray detector.

Since the  scattering angle of gamma-rays can be measured
via reconstruction of the Compton scattering in the Compton camera, the
SGD is capable of measuring polarization of celestial sources brighter
than a few $\times$ 1/100 of the Crab Nebula, polarized 
above $\sim$ 10 \%. This capability is expected to yield polarization measurements 
in several celestial objects, providing new insights into properties of soft gamma-ray
emission processes\cite{Ref:Tajima-pol,Ref:Tajima2010}.

Last year, the technology behind the SGD was proved  in measurements of the distribution of  Cs-137 in the environment
of Fukushima\cite{Ref:Compton_Fukushima}. In addition to showing that the SGD technology works as designed `in the field'' the use of the Si/CdTe Compton Camera provided crucial information in understanding how the fall out from the 2011 nuclear accident was distributed.

 \section{Science Operation}
 
 ASTRO-H is in many ways similar to Suzaku in terms of orbit,
pointing, and tracking capabilities, although the mass is considerably larger (the total mass 
at launch will be 2700~kg, nearly double Suzaku's 1700 kg).
 ASTRO-H will be launched into a circular orbit 
with altitude of 500--600~km, and inclination of 31~degrees.  Science 
operations will be similar to those of Suzaku, with pointed observation of each 
target until the integrated observing time is accumulated, and then slewing 
to the next target. A typical observation will require a few $\times$ 100~ksec integrated 
exposure time. All instruments  are co-aligned and 
will operate simultaneously. The current plan is to use the first three months for check-out and
start the PV phase with observations proprietary to the ASTRO-H team. Guest observing time will 
start from 10 months after the launch. About 75\% of the satellite time will be devoted to
 GO observations after the PV phase is completed. We are planning to implement key-project
type observations in conjunction with the GO observation time.

\section{Expected Scientific Performance}


The spectroscopic capability of X-ray micro-calorimeters is unique in X-ray astronomy, since no other previously or currently operating spectrometers could  achieve comparable high energy resolution, high quantum efficiency, and spectroscopy for spatially extended sources at the same time. Imaging spectroscopy with the Soft X-ray Spectrometer (SXS) of extended sources can reveal line broadening and Doppler shifts due to turbulent or bulk velocities. This capability enables the determination of the level of turbulent pressure support in clusters, SNR ejecta dispersal patterns, the structure of AGN and starburst winds, and the spatially dependent abundance pattern in clusters and elliptical galaxies. The SXS can also measure the optical depths of resonance absorption lines, from which the degree and spatial extent of turbulence can be inferred. Additionally, the SXS can reveal the presence of relatively rare elements in SNRs and other sources through its high sensitivity to low equivalent width emission lines. The low SXS background ensures that the observations of almost all line-rich objects will be photon limited rather than background limited.

The imaging capabilities at high X-ray energies will open a new era in high spatial-resolution studies of astrophysical sources of non-thermal emission above 10 keV, probed simultaneously with lower energy imaging spectroscopy.
This
will enable us to track  the evolution of active galaxies with accretion flows which are
heavily obscured, in order to accurately assess their contribution to
black hole growth over cosmological time scale.   
It will also uniquely allow
mapping of the spatial extent of the hard X-ray emission in diffuse sources,thus tracing the sites of cosmic ray acceleration in structures ranging in size from megaparsecs, such as clusters of galaxies, down to parsecs, such as young supernova remnants\cite{Ref:Koyama,Ref:Uchiyama,Ref:Aharonian}.  
Those studies will be complementary to the SXS measurements:  
observing the hard X-ray synchrotron emission will allow a study of the most energetic 
particles, thus revealing the details of particle acceleration mechanisms 
in supernova remnants, while the high resolution SXS data on the gas kinematics 
of the remnant will constrain the energy input into the accelerators.

\begin{figure}[htbp]
\includegraphics*[width=6cm,angle=90]{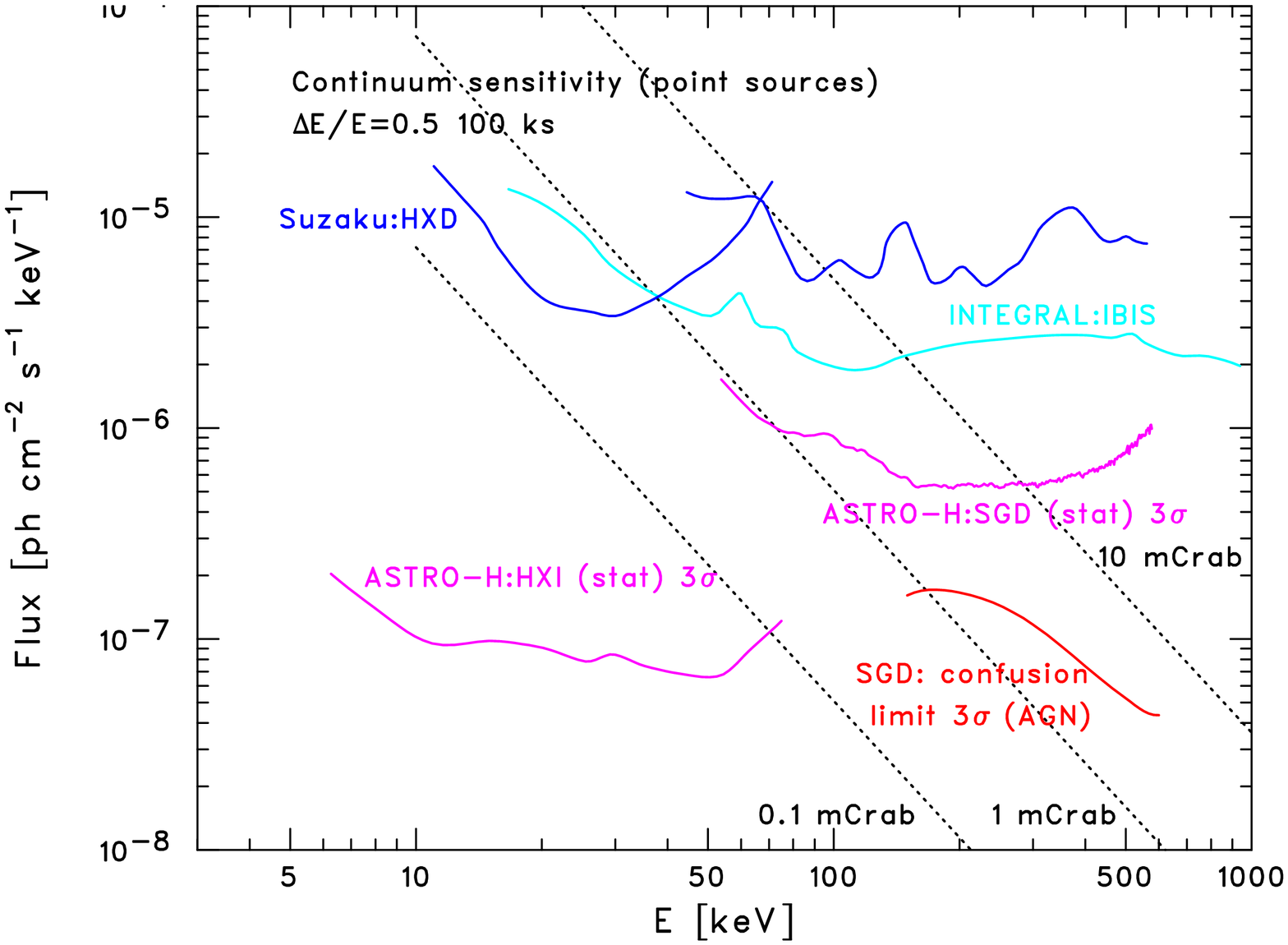}
\includegraphics*[width=6cm,angle=90]{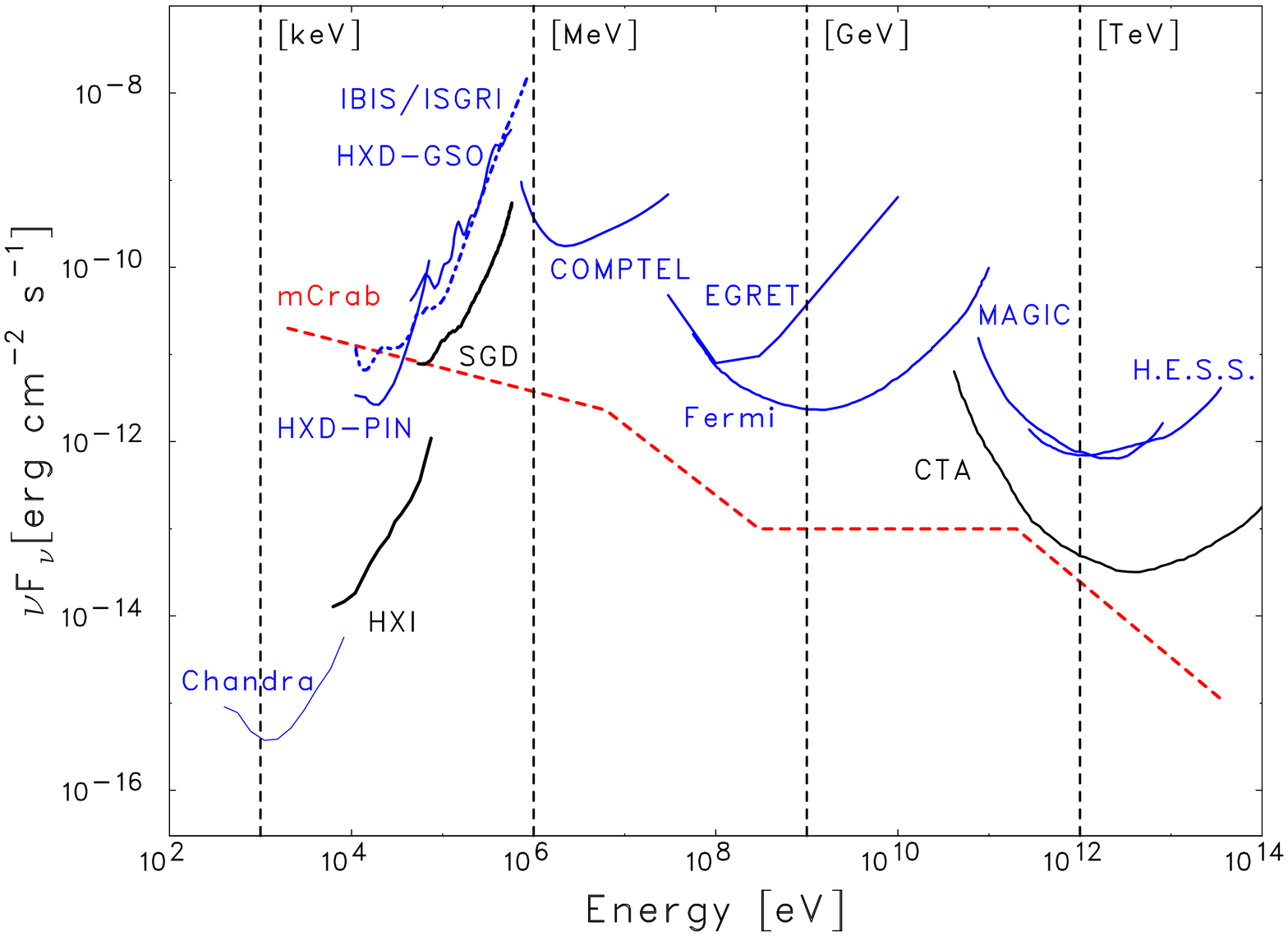}
\caption{(left) The $3\sigma$ sensitivity curves for the HXI and SGD onboard ASTRO-H for an isolated point source. (100 ks exposures and $\Delta E/E = 0.5$) (right) Differential sensitivities of different X-ray and $\gamma$-ray instruments for an isolated point source\cite{Ref:Takahashi_CTA}. 
Lines for the \chandra/ACIS-S, the \suzaku/HXD (PIN and GSO), the \integral/IBIS (from the 2009 IBIS Observer's Manual), and the \astroh/HXI,SGD are the $3\sigma$ sensitivity curves for 100 ks exposures. A spectral bin with $\Delta E/E = 1$ is assumed for \chandra\ and $\Delta E/E = 0.5$ for the other instruments. }
\label{Fig:Sensitivity}
\end{figure}

As shown in Figure\,\ref{Fig:Sensitivity}, the sensitivity to be achieved by \astroh\ (and similarly \nustar) is about two orders of magnitude improved compared to previous collimated or coded mask instruments that have operated in this energy band.
 This will bring a breakthrough in our understanding of hard X-ray spectra of celestial sources in general. With this sensitivity, $30-50$\,\% of the hard X-ray Cosmic Background will be resolved. This will enable us to track the evolution of active galaxies with accretion flows that are heavily obscured, in order to accurately assess their contribution to the Cosmic X-ray Background (i.e., black hole growth)  over cosmic time. In addition, simultaneous observations of blazar-type active galaxies with \lat\ and the TeV $\gamma$-ray telescopes are of vital importance to study particle acceleration in relativistic jets.  An exciting and unique possibility is the detection of the polarization in hard X-rays by SGD during particularly strong flaring states of the brightest BL Lacs, like that of Mrk\,501 in 1997 when the synchrotron continuum of the source extended up to the $\gtrsim 100$\,keV photon energy range\cite{Ref:Takahashi_CTA,Ref:Tajima-pol}.

In the following sections, we give a few examples of key science which will be addressed by ASTRO-H.

\subsection{Supernova Remnants}

The high resolution X-ray spectroscopy provided by ASTRO-H will be particularly 
ground-breaking for supernova remnants (SNRs) because they are extended objects 
with rich emission-line spectra from a wide range of different elements (carbon 
through nickel). To determine the element abundances reliably, measurements of 
the relative strengths of a number of lines from each elemental species are 
required. Accurate element abundances provide constraints to test the explosion 
mechanisms of supernovae and to explore their environments.  Gas motions of 
the rapidly expanding supernova ejecta and swept-up interstellar/circumstellar  
medium can also be measured by ASTRO-H via their Doppler shifts.  
Velocity measurements inferred from such Doppler shifts are needed to understand how SNRs evolve, 
based on their age and the detailed properties of the explosion, the ejecta, 
and ambient medium.
\begin{figure}
\begin{center}
\includegraphics*[scale=0.4]{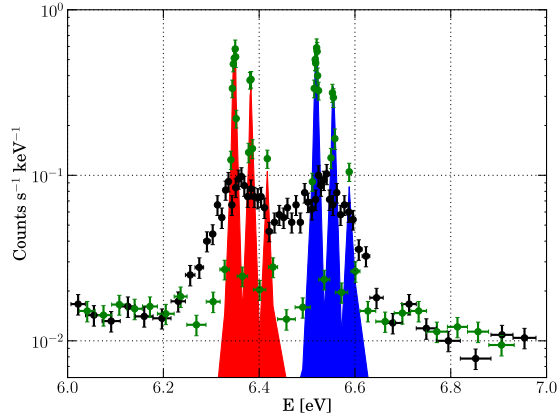}
\end{center}
\caption{
Simulated \astroh\ SXS spectrum (black points) around the iron K-shell complex of Tycho's SNR for an exposure of 100\,ks\cite{Ref:AHQref}. 
For  reference, a simulated spectrum with no 
thermal Doppler broadening is also shown (green points). 
Iron K lines from a blob that is receding are red-shaded, while 
lines from an approaching blob are blue-shaded.  }
\label{fig:Tycho}
\end{figure}

The excellent resolution of the \astroh\ SXS offers an opportunity to measure the temperature of shocked iron ions in young SNR, 
a potential breakthrough in understanding the physics of shock heating. 
Figure~\ref{fig:Tycho} presents simulated spectra of SXS observations of the central portion of Tycho's SNR. 
The
spectrum shown with black points in Fig.~\ref{fig:Tycho} corresponds to the emission
produced by two plasma blobs which were assumed to be receding and approaching us with $\pm 4000\ \rm km\ s^{-1}$  with the same 
parameters\cite{Ref:Hayato}: an iron temperature of $kT_{\rm Fe} = 3$ MeV (mass-proportional heating),  
an electron temperature of $kT_e = 5$ keV,  and an ionization parameter 
of $\tau = 0.9\times 10^{10}$\,s\,cm$^{-3}$. For a reference, a simulated spectrum with no thermal Doppler broadening is also shown (green points)\cite{Ref:Takahashi_CTA}.

\begin{figure}
\centerline{\includegraphics[height=6cm,clip]{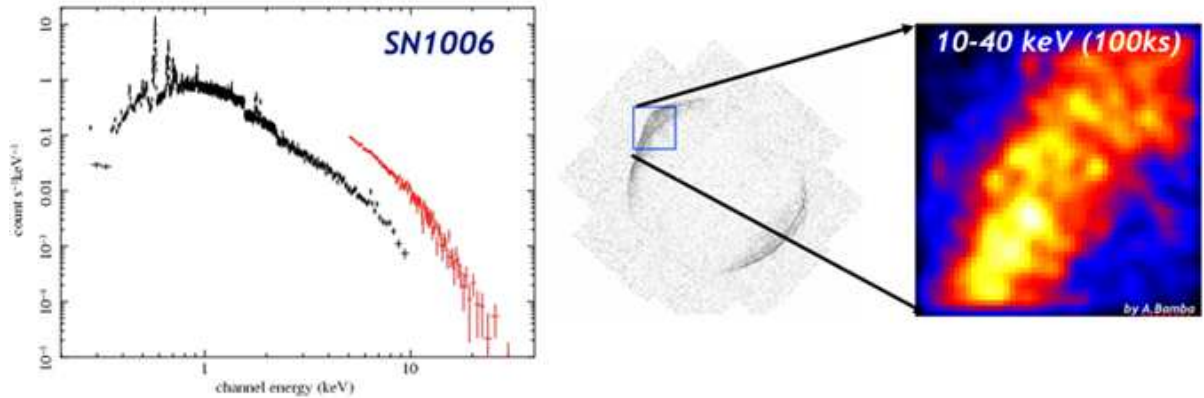}}
\caption{(left)  Simulated spectra for 100 ks SXS observation of SN1006 with SXS and HXI. (right) Image expected from the combination of HXT and HXI.}
\label{Fig:SNR}
\end{figure}

 Particle acceleration is receiving much attention at present, 
but the origin of cosmic rays is still unclear 100 years after their discovery.  
Young SNRs with shock speeds of several 1000 km/s are among the best candidates 
to accelerate cosmic rays up to energies around $10^{15}$ eV (the ``knee'' in the observed cosmic ray spectrum).
 The combination of ASTRO-H's hard X-ray 
imaging capability and high spectral resolution will provide a unique insight into several crucial aspects of shock acceleration in SNRs such as the maximum energy of the accelerated particles, background conditions around the shock fronts, and the particle acceleration efficiency (See Fig.~\ref{Fig:SNR}).
High energy protons in
SNR can be also studied with HXI through the detection of the
synchrotron emission produced by secondary pair generated at
proton-proton collisions.

 \subsection{A Census of Obscured Active Galactic Nuclei}

Recent observations imply the existence of a large number of Active Galactic Nuclei (AGN) that are heavily
obscured by the gas and dust surrounding their supermassive black holes\cite{Ref:Ueda,Ref:Ueda2}. 
 Some of those are identified as a new type of AGN, so deeply buried in dense gas that they show little emission in the soft X-ray or visible/ultraviolet light. While this has made such objects extremely difficult to detect and observe, this highly obscured activity may in fact represent the dominant phase of  the supermassive black hole growth\cite{Ref:Treister}.
Understanding this phase is thus crucial for a proper understanding of the correlated evolution of supermassive black hole and their host galaxies. As shown in Fig. 12, the high sensitivity for hard X-rays provided by HXI allows precise spectral studies of even very obscured AGN. ASTRO-H will provide us with a large AGN sample to pursue systematic studies of the true AGN population, unbiased by obscuration effects and extending to far higher redshifts than current hard X-ray-selected samples\cite{Ref:Ballantyne}, allowing us to measure the co-evolution of supermassive black
holes with their host galaxies.

By assuming a background level of $\sim$ 1$\times$10$^{-4}$ counts/s/cm$^{2}$/keV, in which
the non-X-ray background is dominant, the source detection limit in
1\,Msec in the $10- 80$ keV band is roughly 10$^{-14}$\,erg\,cm$^{-2}$\,s$^{-1}$ 
(for a power-law spectrum with a photon index of 2). This is about two orders 
of magnitude better than previous instrumentation, and thus 
will result in a breakthrough in our understanding of
hard X-ray spectra. With this sensitivity, 30-50~\% of hard X-ray Cosmic
Background  is expected to be resolved\cite{Ref:Ueda} .
In addition to the imaging below 80~keV, the SGD will enable high sensitivity observations in the soft $\gamma$-ray range closely matching the sensitivity of the HXT/HXI combination.
The extremely low background observations 
allowed by the new concept of a narrow-FOV Compton telescope adopted for
the SGD will provide sensitive $\gamma$-ray spectra up to 600~keV, with moderate
sensitivity for polarization measurements. 

\begin{figure}
\centerline{\includegraphics[height=7cm,clip]{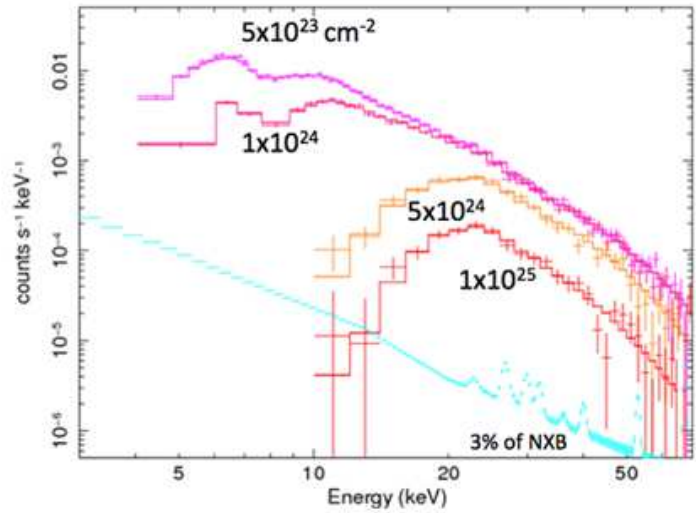}}
\caption{Simulated HXI spectra of heavily obscured AGN with different  absorbing columns (N$_{\rm{H}}$) for an exposure of 100 ks.  The continuum is assumed to be  a power-law of photon index 1.9, with an intrinsic 2-10 keV flux of 1x10$^{-11}$ erg/cm$^{2}/$s  (based on Swift J0601.9-8636\cite{Ref:Ueda2}). The scattered component and Fe lines are not included.
The light blue line represents the uncertainty in the background for a point source.}
\label{Fig:BH}
\end{figure}

\subsection{Black Hole Vicinities}

ASTRO-H provides an unprecedented view of the motions and extreme physical conditions of matter near the event horizon. 
This will help us to unravel how black holes grow by accreting matter and simultaneously how they influence their environments via radiative and mechanical energy output in the form of the intense emission fields and powerful, often relativistic outflows of magnetized plasma that accompany the accretion process.

\begin{figure}
\centerline{\includegraphics[height=7cm,clip]{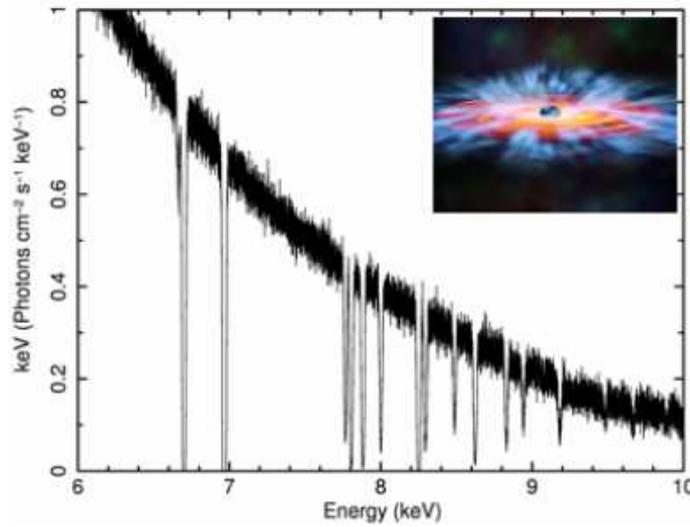}}
\caption{Simulated spectrum of the Galactic black hole candidate GRO J1655-40 for a 100 ks SXS observation\cite{Ref:AHQref}, based on a model that explains a prior Chandra observation.The dips in the spectrum are due to absorption from a highly ionized outflowing wind, comprised of atomic species from oxygen to nickel, that is driven by magnetic fields\cite{Ref:Miller}.  }
\label{Fig:Wind}
\end{figure}

Gaseous winds driven from black hole disks can carry away a substantial fraction of the gas that would otherwise accrete onto the central black hole. Such winds are hot, and most easily detected in the Fe K band via absorption lines. As shown in Fig. \ref{Fig:Wind}, the superior resolution of SXS in the Fe K band enables the unambiguous detection of weak and narrow lines from a wind. We will be able to use these to precisely determine the radius at which the wind is launched and the mass outflow rate carried by it. This will give us strong constraints on the driving mechanism of the wind and its feedback on the accretion flow as well as the black hole's environment\cite{Ref:Fabian}.

On the smallest scales, many active galactic nuclei (AGN) show signatures 
from the the innermost parts of their accretion disk in the form of broad ``relativistic" Fe K 
emission lines. These broad lines were discovered using ASCA in the early 1990s 
and have been confirmed by XMM-Newton and Suzaku\cite{Ref:YTanaka,Ref:Reeves}. 
There is, however, a complex relationship between the Fe K line properties, the 
underlying continuum, and the signatures of cold and/or
partially ionized material near the AGN. Precise measurements of the complex 
Fe K line and absorption components require high spectral resolution.  
The measurement of changes in the X-ray emission and absorption spectral 
features, on the orbital time  scale of black holes in AGN, will 
enable characterization of the velocity field and ionization state of the 
plasma closest to the event horizon.  The optically thick material that produces 
the broad fluorescent Fe K line also creates a Compton reflection ``hump'' peaking 
at $E>20$~keV detectable with hard X-ray and soft gamma-ray detectors, 
providing multiple insights into the physics of the disk. In order to understand 
the evolution of environments surrounding supermassive black holes, 
high signal-to-noise measurements of the broad lines of many  AGN are needed 
up to, at least, $z$$\sim$1. This requires high spectral resolution and bandpass 
extending to at least $\sim$40~keV.  These observations will provide the first 
unbiased survey of broad Fe K line properties  to $z$$\sim$1.
We will sample a few high-z, high-luminosity AGN with adequate S/N and time resolution.

XMM-Newton and Suzaku spectra frequently show time-variable absorption 
and emission features in the 5--10~keV band. If these features are due 
to Fe, they represent gas moving at very high velocities with both red-  
and blue-shifted components from material presumably near the event horizon. 
CCD resolution is insufficient, and the required grating exposures are 
too long, to properly characterize the velocity field and ionization of 
this gas and determine whether it is from close to the black hole or 
from high velocity winds. SXS, in combination with  HXI, provides a 
dramatic increase in sensitivity over Suzaku, enabling measurements 
that probe the geometry of the central regions of $\sim$50 AGNs on 
the orbital timescale of the Fe producing region (for an AGN with 
a 3$\times$10$^{7}M_{\odot}$ black hole, this is  $\sim$60~$GM/c^{3}$ = 10~ksec). 

\subsection{Clusters of Galaxies}

\begin{figure}
\centerline{\includegraphics[height=7cm,clip]{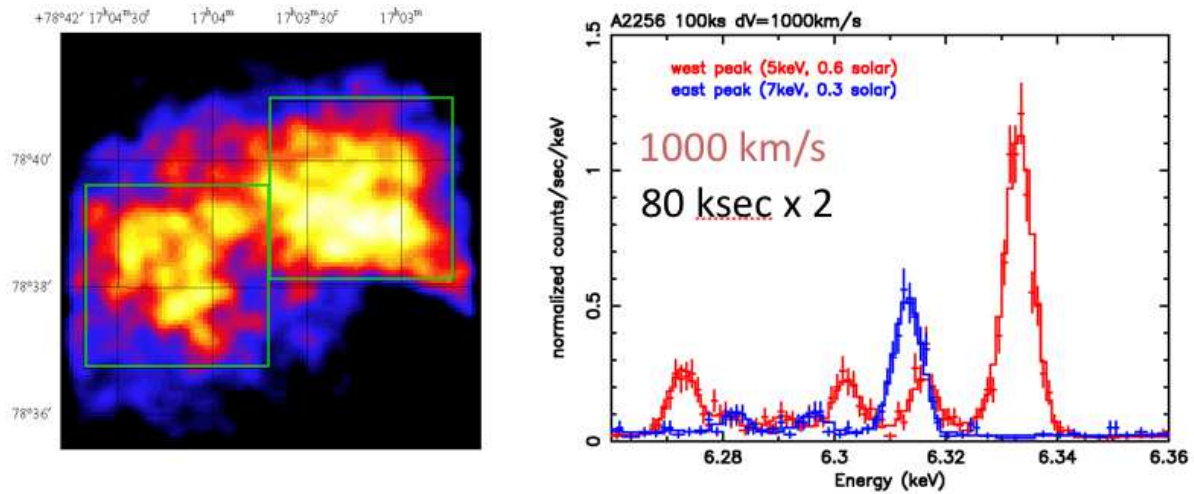}}
\caption{ Simulated image and spectra of a merging cluster A2256 by assuming
1000 km/s difference of line of sight speed of hot gas in the cluster. The simulation is for 80~ksec each.}
\label{Fig:SXS1}
\end{figure}

\begin{figure}
\centerline{\includegraphics[height=5.5cm,angle=0]{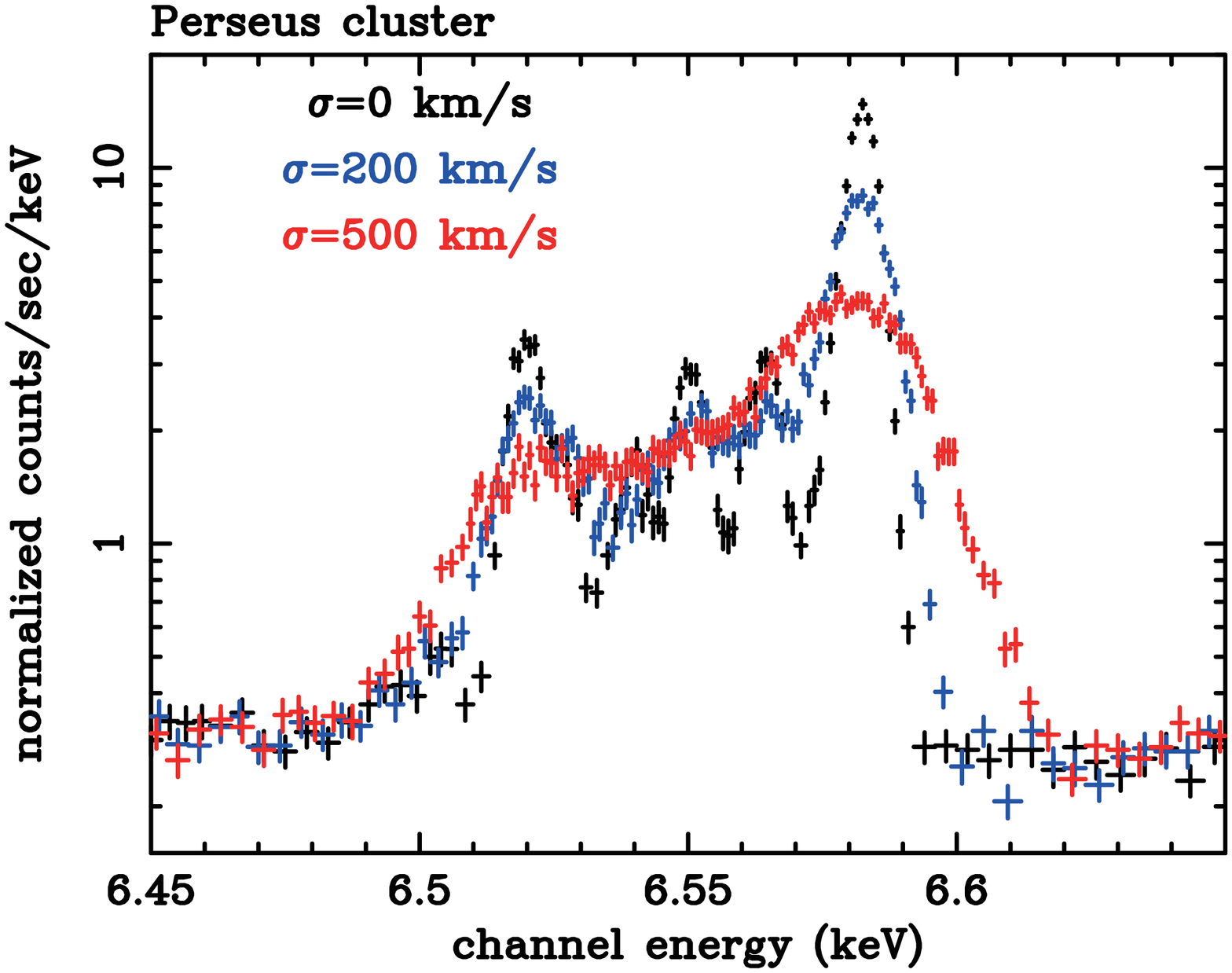}\hspace{5mm}
\includegraphics[height=5.5cm,angle=0]{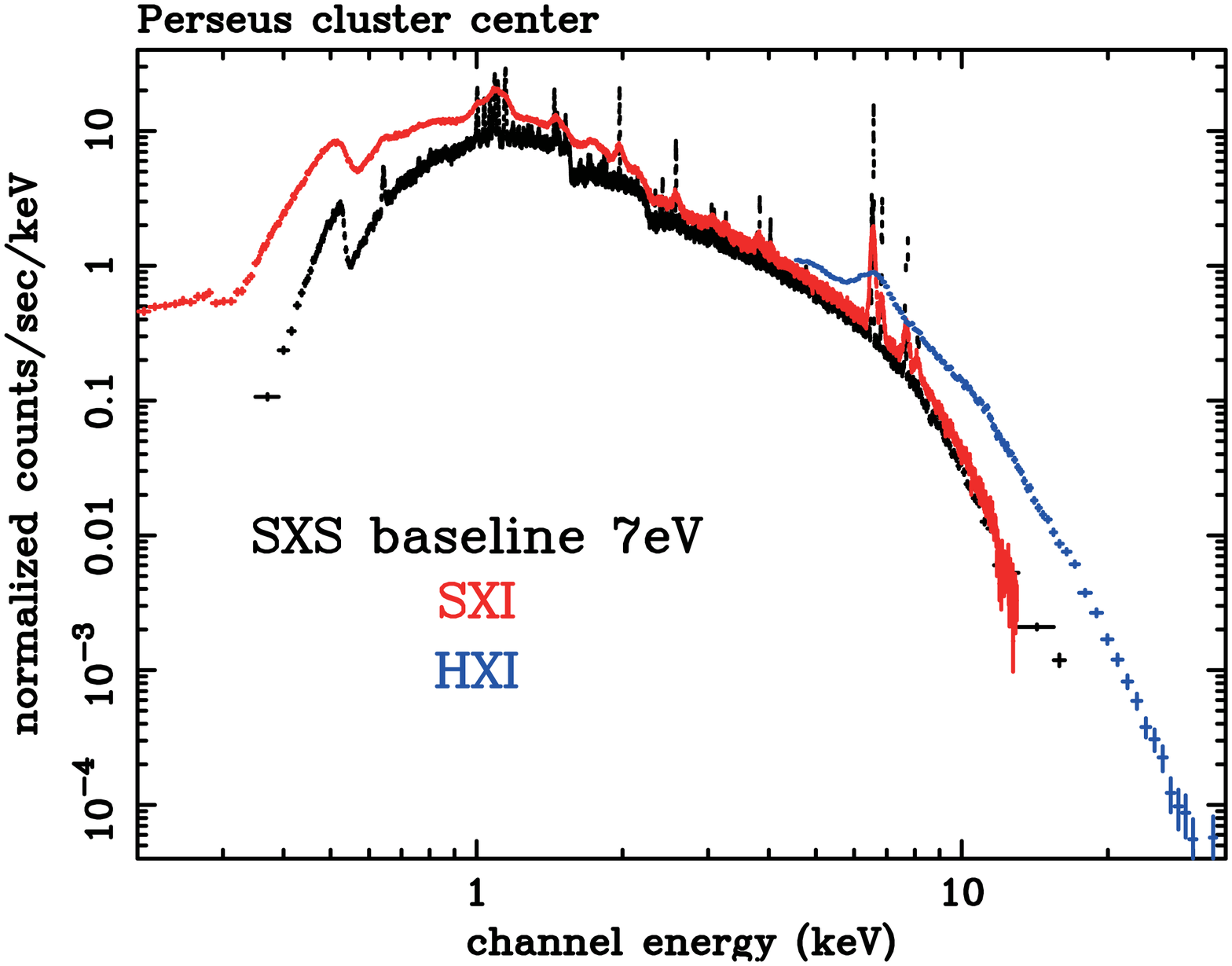}}
\caption{Simulated spectra for 100\,ks \astroh\ observations of Perseus Cluster. {\bf (left)} 
SXS spectra around the iron K line complex.  
Line profiles assuming $\sigma=0$, 200 and 500\,$\rm km\ s^{-1}$ turbulence. 
{\bf (right)} SXS (black), SXI (red), and HXI (blue) spectra 
for hot plasma with a mixture of three different temperatures of 0.6, 2.6 and 6.1\,keV ($r < 2'$)\cite{Ref:Takahashi2010}.}
\label{Fig:SXS2}\end{figure}

All studies of the total energy content (including that of non-thermal 
particles), aimed to draw a more complete picture of the
high energy universe, require observations by {\sl both} a spectrometer 
capable of measuring the bulk plasma velocities and/or turbulence with the 
resolution corresponding to the speed of a few $\times$ 100~km/s {\sl and}
an arc-min imaging system in the hard X-ray band, with  sensitivity two-orders of 
magnitude better than previous missions (See Fig.~\ref{Fig:SXS1} 
and Fig.~\ref{Fig:SXS2}).
In clusters, X-ray  emitting hot gas is trapped in the gravitational potential
well, and shocks and/or turbulence are produced in this gas, as smaller
substructures with their own hot gas halos fall into and merge with the
dominant cluster. Large scale shocks can also be produced as gas from the
intracluster medium falls into the gravitational potential of a cluster.

There is a strong synergy between the hard X-ray imaging data and the high resolution ($\Delta E $ $\leqq$ 7~eV) soft X-ray spectrometer: the kinematics of the gas, probed by the width and energy of the emission lines, constrains the energetics of the system,. 
The  kinematics of the gas provides information about the bulk motion; the energy of this motion is in turn
responsible for acceleration of particles to very high energies at
shocks, which is in turn manifested via non-thermal emission processes, best
studied via sensitive hard X-ray measurements.

Precision cosmology uses astronomical observations to determine the 
large-scale structure and content of the Universe. Studies of clusters 
of galaxies have provided independent measurements of the dark energy 
equation of state and strong evidence for the existence of dark matter. 
Using a variety of techniques (including the growth of structure, the 
baryonic fraction in clusters, and the Sunyaev-Zel'dovich effect) 
a well-constructed survey of clusters of galaxies, with the necessary 
supporting data\cite{Ref:Rapetti}, can provide precise measurements of 
cosmological parameters, including the amount and properties of dark energy 
and dark matter. The key step is to connect observables (such as flux and 
temperature) to cluster masses. 
Currently, large area cluster surveys are being carried out using the Sunyaev-Zel'dovich Effect by the Atacama Cosmology Telescope, Planck, and the South Pole Telescope, to be joined in the near future by the eROSITA X-ray mission.
To reduce the systematic 
uncertainties on the masses inferred from the coarser data from these surveys, 
a training set of precise cluster masses must be obtained. Measurements of the bulk 
motion of clusters of galaxies and amounts of non-thermal energies going 
to cosmic-ray acceleration could reduce the ``{\dots} substantial uncertainties 
in the baryonic physics which prevents their use at a high level of precision at 
the present time" (Dark Energy Task Force\cite{Ref:Albrecht}). Line diagnostics 
with energy resolution of $\Delta E $ $\leqq$ 7~eV greatly reduce the uncertainties in the baryonic 
physics by determining the velocity field, any deviations from thermal 
equilibrium, and an accurate temperature for each cluster. Information 
about the non-thermal particle content of clusters can be determined 
via measurements of their Compton upscattering of the CMB:  this is best 
studied via hard X-ray imaging, providing additional clues about the physical 
state of the cluster gas.  

\section{Summary}

The ASTRO-H mission objectives are: to determine the evolution of yet-unknown 
obscured supermassive black holes (SMBHs) in Active Galactic Nuclei (AGN); 
to trace the growth history of the largest structures in the Universe;
to trace the chemical evolution of the universe;
to probe feedback from the growth of supermassive black holes onto their galaxy and cluster environments;
to provide insights into the behavior of material in extreme gravitational fields; 
to determine the spin of black holes and the equation of state of neutron stars; 
to trace particle acceleration structures in clusters of galaxies and SNRs; 
and to investigate the detailed physics of astrophysical jets. 

ASTRO-H will open a completely new field of spatial studies of non-thermal emission across a broad range of energies extending well above 10 keV with hard X-ray telescopes and  enable us to track the evolution of active galaxies with accretion flows that are heavily obscured. It will also uniquely allow mapping of the spatial extent of the hard X-ray emission in diffuse sources, thus tracing the sites of particle acceleration in structures ranging in size from clusters of galaxies down to supernova remnants.
At even higher energies, the sensitivity of the SGD will allow us for the first time to routinely detect and characterize the spectra of bright AGN above 100 keV.

The key properties of SXS onboard ASTRO-H are its high spectral resolution for both 
point and diffuse sources over a broad bandpass ($\leq$7~eV FWHM throughout 
the 0.3--12~keV band), high sensitivity (effective area of 160~cm$^2$ at 1~keV 
and 210~cm$^2$ at 7~keV), and low non-X-ray background 
(1.5$\times$10$^{-3}$~cts~s$^{-1}$keV$^{-1}$). These properties open up 
a full range of plasma diagnostics and kinematic studies of X-ray emitting 
gas for thousands of targets, both Galactic and extragalactic. SXS improves 
upon and complements the current generation of X-ray missions, including 
Chandra, XMM-Newton, Suzaku, Swift and NuSTAR. 

\section*{Acknowledgments}

The authors are deeply grateful for on-going contributions provided by other members in the ASTRO-H team in Japan, the US, Europe and Canada.

\end{document}